\documentclass[apj,square,numbers]{emulateapj}
\usepackage{graphicx}
\usepackage{latexsym}
\usepackage{epsfig}
\usepackage{color}
\usepackage[colorlinks, linkcolor=blue, citecolor=blue, urlcolor=blue]{hyperref}
\usepackage{amsmath,amssymb,bm,color}
\usepackage{ulem}
\def\bea{\begin{eqnarray}}
\def\eea{\end{eqnarray}}
\def\be{\begin{equation}}
\def\ee{\end{equation}}

\newcommand{\de}{\mathrm d}

\newcommand{\g}{$\gamma$}

\shorttitle{Cross-correlation of $\gamma$ rays with cluster catalogs}
\shortauthors{Branchini et al.}

\begin{document}
\title{Cross-correlating the $\gamma$-ray sky with catalogs of galaxy clusters}

\author{Enzo Branchini$^{2,3,4}$}
\author{Stefano Camera$^{5}$}
\author{Alessandro Cuoco$^{6}$}
\author{Nicolao Fornengo$^{7,8}$}
\author{Marco Regis$^{7,8}$}
\author{Matteo Viel$^{9,10,11}$}
\author{Jun-Qing Xia$^1$}

\affiliation{$^1$ Department of Astronomy, Beijing Normal University, Beijing 100875, China}
\affiliation{$^2$ Dipartimento di Matematica e  Fisica, Universit\`a degli Studi ``Roma Tre'', via della Vasca Navale 84, I-00146 Roma, Italy}
\affiliation{$^3$ INFN, Sezione di Roma Tre, via della Vasca Navale 84, I-00146 Roma, Italy}
\affiliation{$^4$ INAF Osservatorio Astronomico di Roma, Osservatorio Astronomico di Roma, Monte Porzio Catone, Italy}
\affiliation{$^5$ Jodrell Bank Centre for Astrophysics, The University of Manchester,\\Alan Turing Building, Oxford Road, Manchester M13 9PL,UK}
\affiliation{$^6$ Institute for Theoretical Particle Physics and Cosmology, RWTH Aachen University, 52056 Aachen, Germany}
\affiliation{$^7$ Dipartimento di Fisica, Universit\`{a} di Torino, via P. Giuria 1, I--10125 Torino, Italy}
\affiliation{$^8$ Istituto Nazionale di Fisica Nucleare, Sezione di Torino, via P. Giuria 1, I--10125 Torino, Italy}
\affiliation{$^9$ INAF Osservatorio Astronomico di Trieste, Via G. B. Tiepolo 11, I-34141, Trieste, Italy}
\affiliation{$^{10}$ SISSA-International School for Advanced Studies, Via Bonomea 265, 34136 Trieste, Italy}
\affiliation{$^{11}$ INFN, Sezione di Trieste, via Valerio 2, I-34127, Trieste, Italy}

\email{For correspondence: regis@to.infn.it}


\begin{abstract}
We report the detection of a cross-correlation signal between {\it Fermi} Large Area Telescope  diffuse \g-ray maps and catalogs of clusters.
In our analysis, we considered three different catalogs: WHL12, redMaPPer and PlanckSZ.
They all show a positive correlation with different amplitudes, related to the average mass of the objects in each catalog, which also sets the catalog bias.
The signal {detection} is {confirmed} by {the results of} a stacking analysis.
The cross-correlation signal extends to rather large angular scales, around 1 degree, that correspond, at the typical redshift of the clusters in these catalogs,
to a few to tens of Mpc, i.e. the typical scale-length of the large scale structures in the Universe.
Most likely this signal is contributed by the cumulative emission from AGNs associated to the filamentary structures that converge toward  the high peaks of the matter density field in which galaxy clusters reside. 
In addition, our analysis reveals the presence of a second component, more compact in size and compatible with a point-like emission from within individual clusters.
At present, we cannot distinguish between the two most likely interpretations for such a signal, i.e. whether it is produced by AGNs inside clusters or if it is a diffuse \g-ray emission from the intra-cluster medium.
We argue that this latter, intriguing, hypothesis might be tested by applying this technique to a low redshift large mass cluster sample.

\end{abstract}


\maketitle

\section{Introduction}
\label{sec:intro}

Galaxy clusters are the largest virialized objects in the Universe formed by the gravitational instability-driven 
hierarchical structure formation process. They  are also unique astrophysical laboratories hosting galaxies, highly 
ionised hot gas in thermal equilibrium, dark matter (DM)  and a population of relativistic cosmic rays (CRs). 
The last two components can provide the conditions in which a diffuse \g-ray emission can be produced. 
CRs can lead to the the emission of \g-ray photons via three channels: inverse Compton 
of relativistic electrons with the cosmic microwave background, non-thermal bremsstrahlung, and decay of $\pi^0$ produced through 
collision of relativistic protons with thermal protons (see, e.g., \cite{2010MNRAS.409..449P} and \cite{2016MNRAS.459...70V} for recent simulations of \g-ray emission in galaxy clusters). 
DM can also directly or indirectly produce $\gamma$ rays through 
annihilation or decay, and clusters are promising targets in the particle DM search, due to their large DM content. 

{Clusters of galaxies are not isolated objects. They are located at the node of a complex cosmic web, surrounded by 
a network of filamentary structures populated by astrophysical sources, like the AGNs, that can contribute to the \g-ray emission also from within
the cluster itself.}

The discovery and characterization of cluster-wide $\gamma$-ray emission is therefore important in several ways.
If the signal is induced by CRs, then it can be used to discriminate between different models for the observed radio emission
and clarify the nature of radio halos (see, e.g., \cite{2008SSRv..134...93F}, \cite{2012A&ARv..20...54F} and \cite{2014IJMPD..2330007B} for recent reviews on extended radio emissions in galaxy clusters).
A detection of a signal produced by DM would be its final discovery. In this case, a detailed estimate and understanding of the  contribution of all astrophysical sources to the cluster \g-ray emission is fundamental to unambiguously detect this more exotic signal.

For these reasons, clusters of galaxies have been primary targets for $\gamma$-ray observatories. Yet,
despite numerous efforts, unambiguous detection of extended $\gamma$-ray signal from the intra-cluster medium is lacking.
Upper limits on the emission from individual galaxy clusters have been obtained from the analysis of space-based observations, including the 
EGRET data \citep{2003ApJ...588..155R} and, subsequently, the first 18 months of {\it Fermi} Large Area Telescope (LAT) data \citep{2010ApJ...717L..71A}, and from ground-based observations in the energy band above 100 GeV
(for a complete list of references, see the Introduction in \cite{Ahnen:2016qkt}).
The lack of detection has paved the way for the stacking approach that has been adopted 
in the analyses of the most recent {\it Fermi}-LAT data releases \citep{Zimmer:2011vy}.

In \cite{Dutson:2012ra}, they have stacked $\gamma$-ray data at positions taken from an X-ray flux-limited sample of clusters, further selecting objects with a core-dominated brightest cluster galaxy with high radio flux. \cite{Huber:2013cia} have performed a stacking considering 53 clusters selected from the HIFLUGCS catalog~\citep{2002ApJ...567..716R}. In \cite{Griffin:2014bra}, they have analyzed 78 richest nearby clusters in the Two Micron All-Sky Survey cluster catalog. 
These searches found no evidence for a signal in the stacked data.

\cite{2014A&A...567A..93P} have analyzed 52.5-month {\it Fermi}-LAT data at the positions of the
55 X-ray galaxy clusters from the HIFLUGCS sample.  
Only the brightest objects have been considered in the analysis to maximise the chance of detecting signatures of the neutral pion decay
which should scale with the X-ray flux. 
An excess has been detected with a statistical significance of  4.3 $\sigma$  within a radius of $\sim 0.25$ deg. However, several arguments 
suggest that the signal is mainly produced by AGN, with no evidence of a contribution from the intra-cluster material.

\cite{2014ApJ...787...18A} have similarly searched for  a spatially extended $\gamma$-ray emission
at the locations of 50 HIFLUGS X-ray galaxy clusters in the 4-year {\it Fermi}-LAT data, employing an improved LAT data selection (P7REP).
They have detected a 2.7 $\sigma$ significant excess in a joint-likelihood analysis of  stacked data. This signal, however, 
seems to be {produced by} three objects, Abell 400, Abell 1367 and Abell 311, and has been conservatively attributed to individual sources
(radio galaxies) within the cluster rather than to genuine diffuse emission.

The $\gamma$-ray spectra of galaxy clusters have also been searched for monochromatic
$\gamma$-ray features, but this characteristic signal has not been revealed in the joint likelihood analysis of different samples of clusters \citep{2016JCAP...02..026A,Adams:2016alz}. 
These results have been used to place upper limits on the velocity-averaged DM cross section for self-annihilation into \g\ rays. 

More specialized analyses  have targeted nearby, individual objects
such as the Coma and the Virgo clusters
\citep{2016ApJ...819..149A, 2015ApJ...812..159A,2014MNRAS.440..663Z}.
The analysis of the Coma cluster in \cite{2016ApJ...819..149A} has revealed
an excess emission within the cluster virial radius.  However, 
its statistical significance is  well below the threshold to
claim detection of $\gamma$-ray produced by CRs
interactions in the cluster.
The analysis of the Virgo cluster in \cite{2015ApJ...812..159A} was mainly aimed at indirect DM detection. It has revealed an extended emission  
within a radius of 3 deg, which has been regarded as an artefact due to the incompleteness of the  
interstellar emission model. These results were used to set an upper limit on the $\gamma$-ray flux produced by
both CRs interaction with the intra-cluster medium and DM annihilation.

The goal of this work is to expand the galaxy clusters analyses described above along three directions.  
Firstly we consider three different, non X-ray selected galaxy cluster samples.  Namely: {\it i)} the 
redMaPPer catalog consisting of clusters identified through the red-sequence matched-filter Probabilistic  Percolation cluster finder applied
 to the Sloan Digital Sky Survey  (SDSS) DR8 \citep{2014ApJ...785..104R}, {\it ii)} the WHL12 catalog 
\citep{2012ApJS..199...34W, 2015ApJ...807..178W} obtained from the SDSS DR12 \citep{2015ApJS..219...12A} and {\it iii)}
the Planck catalog of Sunyaev-Zeldovich (PlanckSZ) clusters \citep{2015arXiv150201598P}.
Secondly, instead of stacking signals, we cross-correlate cluster positions in the three catalogs with {\it Fermi}-LAT data
{and compute 2-point statistics } both in configuration and Fourier space. Combining the information from the 2-point angular cross-correlation function 
(CCF) and the cross angular power spectrum (CAPS) allows one to reduce the impact of systematic errors that may 
affect the stacking analysis which, in any case, we also perform to corroborate our results.
We will follow \cite{xia11} and \cite{Xia:2014} for the cross-correlation measurements.
Thirdly, we shall interpret the detected signal in the framework of the halo model along the line pursued in \citep{Regis:2015zka,2015ApJS..221...29C} (see, e.g., \cite{Cooray:2002dia} for a review).

The paper is organized as follows.
Section~\ref{sec:data} describes the {\it Fermi}-LAT data and the catalogs of clusters employed in the analysis.
The measurement of the angular cross-correlation and of the stacked profiles is presented in Section~\ref{sec:meas}.
Section~\ref{sec:models} introduces the theoretical models which are then compared to data in Section~\ref{sec:res}.
We draw our conclusions in Section~\ref{sec:concl}.

\begin{figure}[t]
\centering
\includegraphics[width=0.4\textwidth]{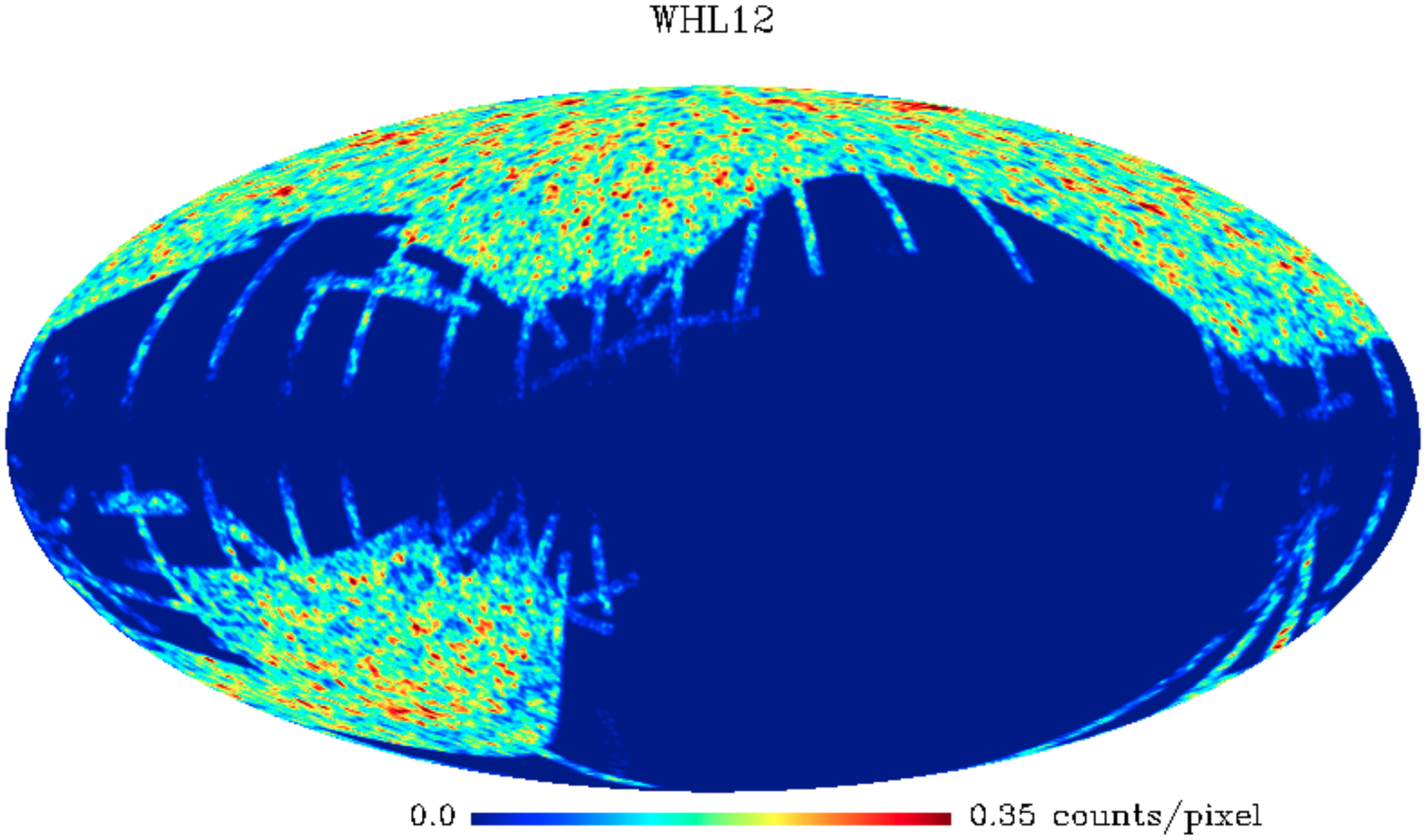}
\includegraphics[width=0.4\textwidth]{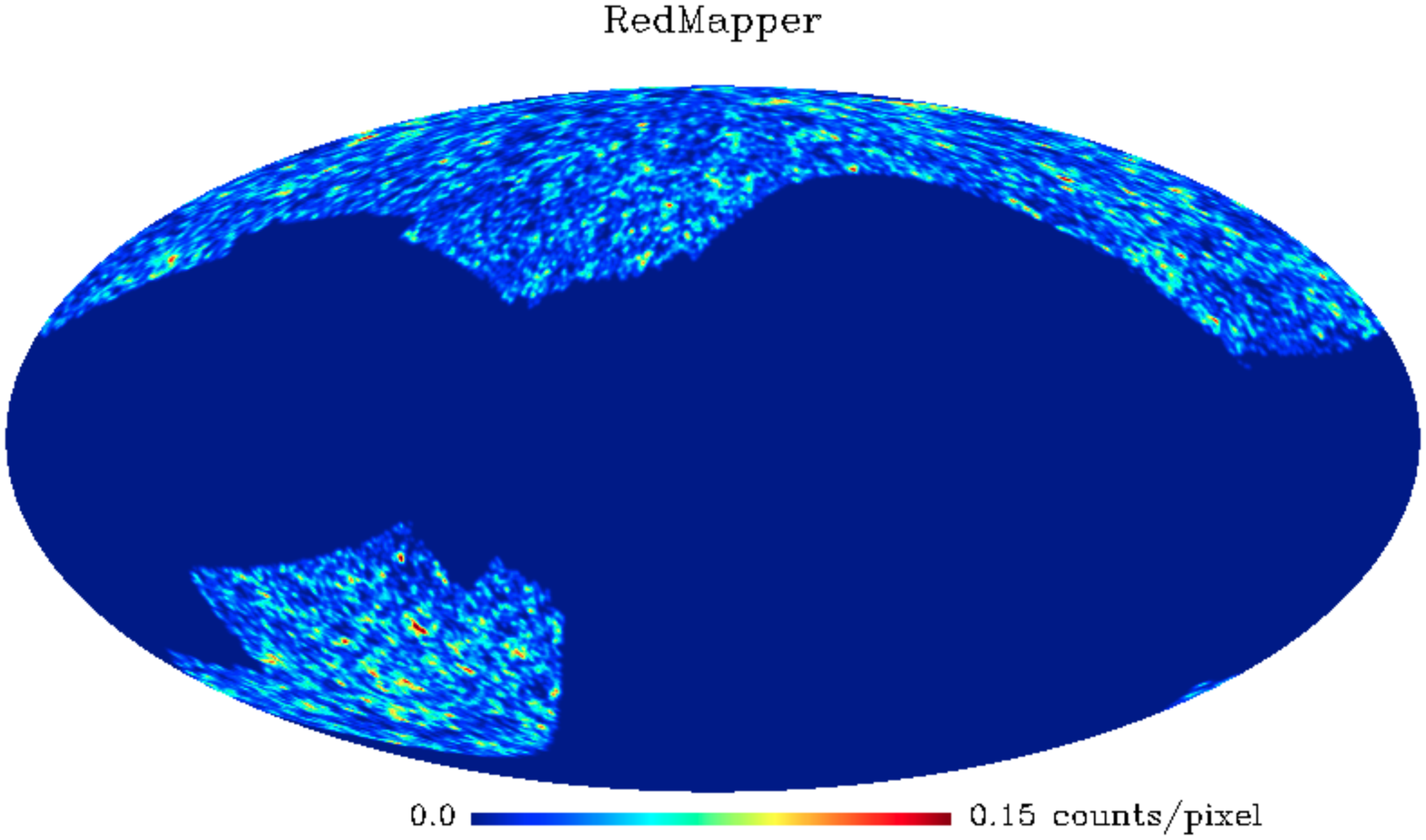}
\includegraphics[width=0.4\textwidth]{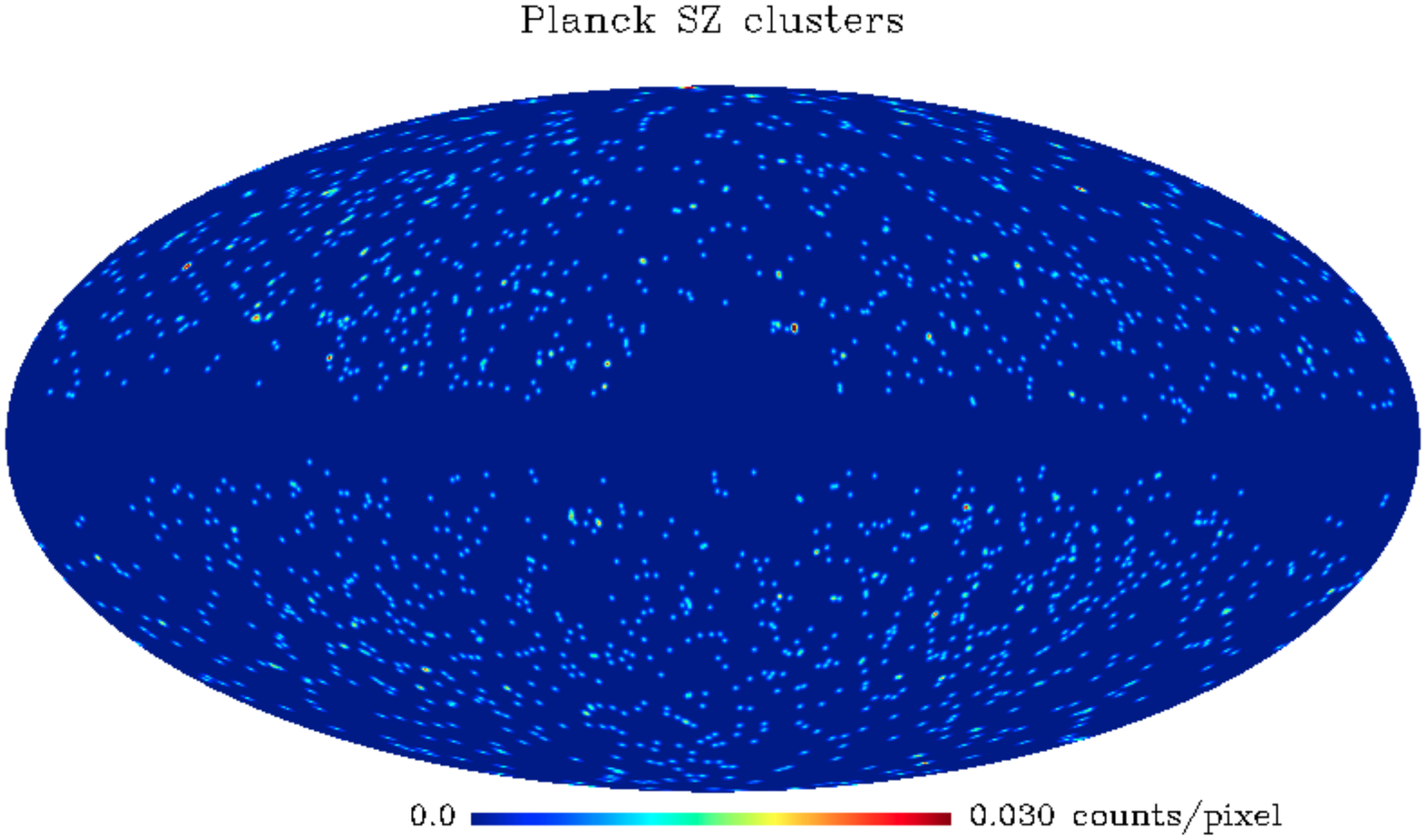}
\caption{Sky-maps of the cluster counts per pixel for the WHL12 (upper panel), redMaPPer (central panel), and PlanckSZ (lower panel) catalogs of clusters. The images have been smoothed to a resolution of $1^\circ$ to improve the visualization.
The maps are shown in Mollweide projection and in Galactic coordinates.
}
\label{fig:maps}
\end{figure}

\section{Data}
\label{sec:data}

In this work we use four different datasets: measurements of high-latitude  diffuse \g-ray emission, and  
three different galaxy cluster catalogs. All of them are described below.

\begin{figure*}[t]
\vspace{-2cm}
\centering
\includegraphics[width=0.33\textwidth]{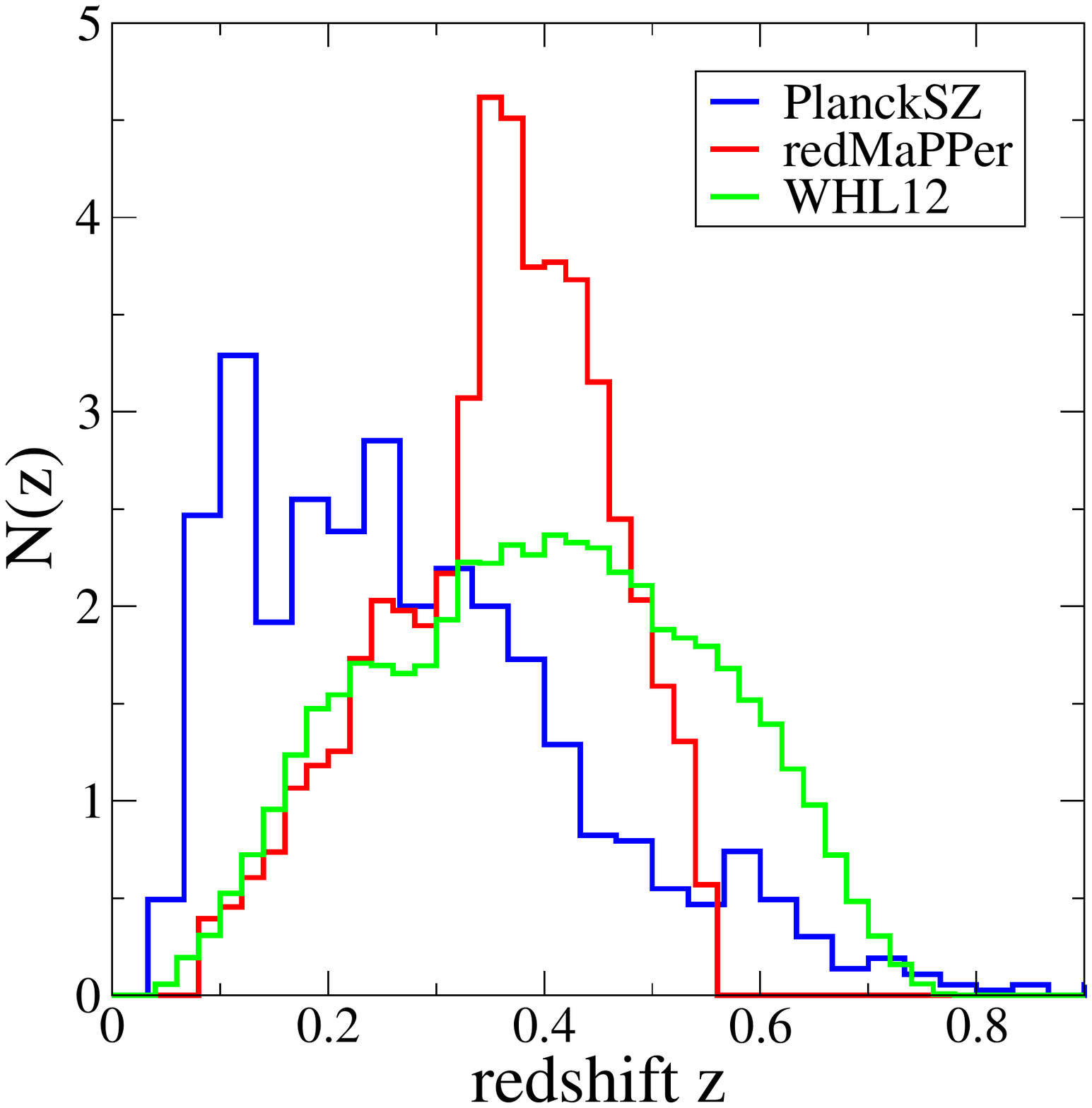}
\includegraphics[width=0.33\textwidth]{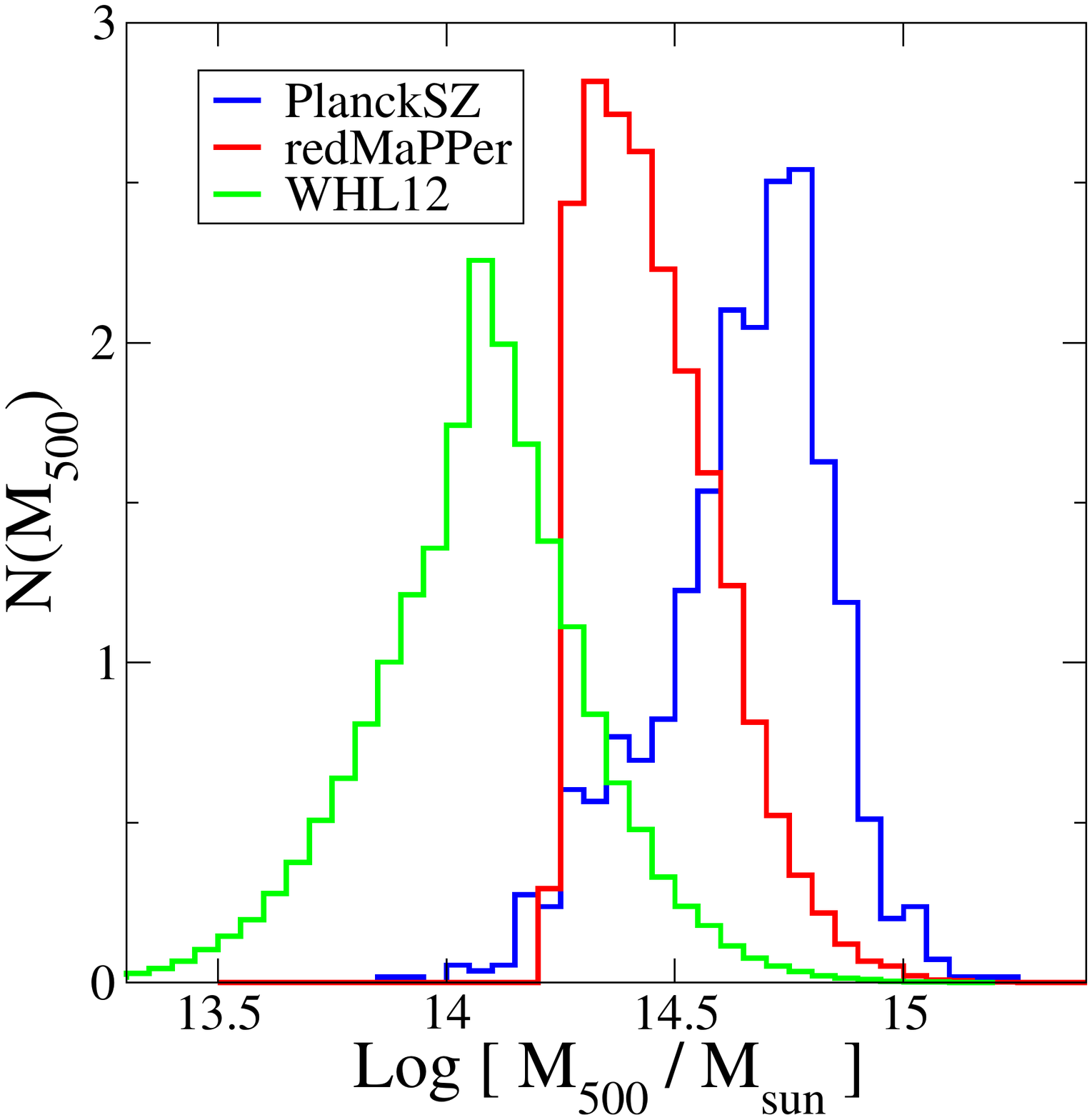}
\includegraphics[width=0.33\textwidth]{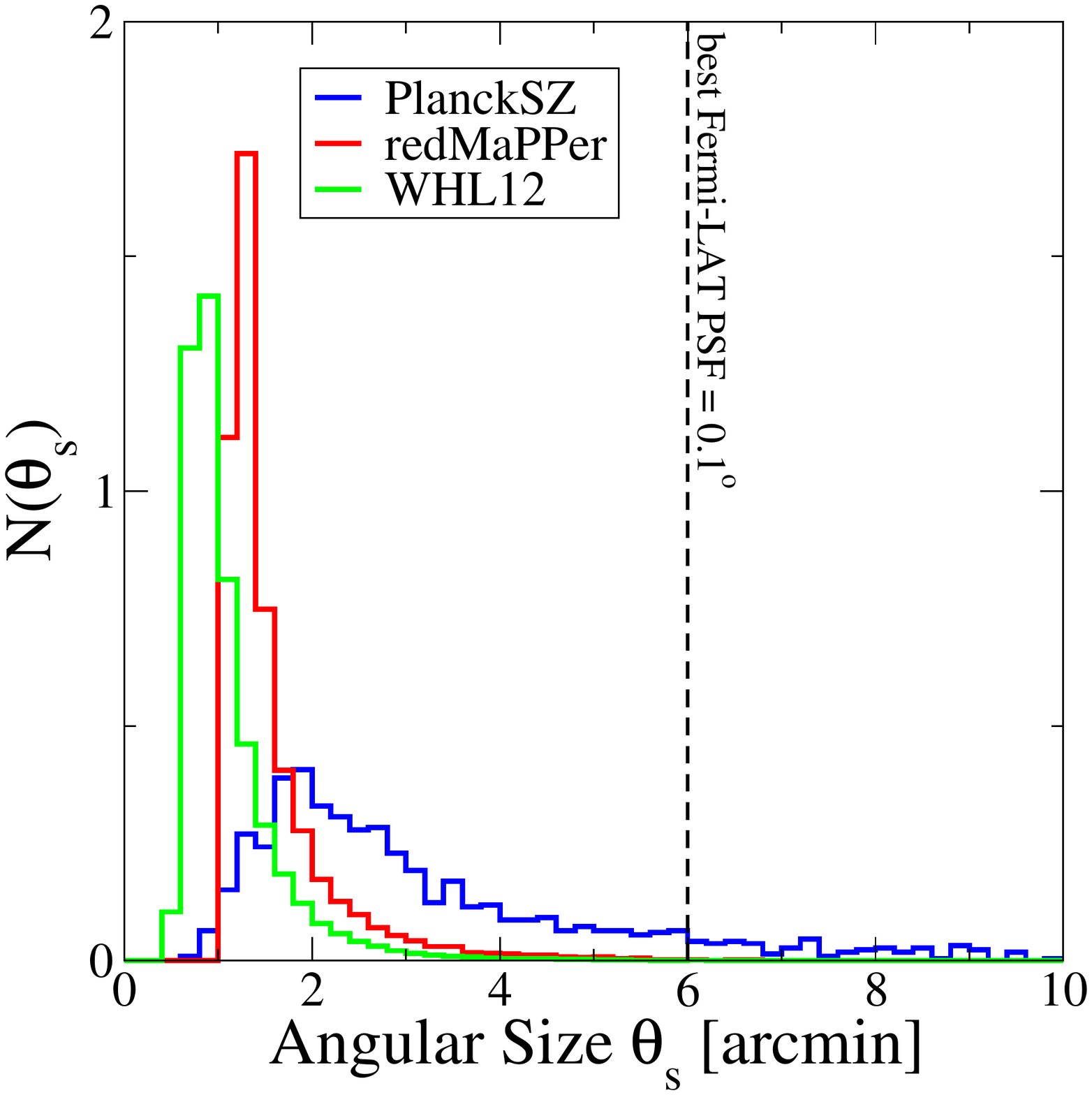}
\caption{Distributions of redshift (left), mass (central) and angular size (right) of the clusters considered in this work. Each histogram, obtained from the catalogs listed in the insets, is normalized to unity. The mass of WHL12 and redMaPPer objects is derived from the reported richness and applying the mass-richness relation in \cite{2015MNRAS.454.2305S} (redMaPPer) and  \cite{2015ApJ...807..178W} (WHL12). For the PlanckSZ objects, we use the mass estimate present in the catalog. The angular size shown in the right panel corresponds to the scale radius of the cluster, namely $\theta_s=r_s/d_A$, where $d_A$ is the angular diameter distance and $r_s=R_{500}/c_{500}$ with $R_{500}=[M_{500}/(4/3\,\pi\,500\,\bar\rho_m(z))]^{1/3}$ and $c_{500}$ being the concentration parameter \citep{Prada:2011jf}.}
\label{fig:cat_prop}
\end{figure*}

\subsection{Fermi-LAT}
\label{sec:fermidata}

{\it Fermi}-LAT is the primary instrument onboard the {\it Fermi} Gamma-ray Space
Telescope launched in June 2008 \citep{2009ApJ...697.1071A}.
It is a \g-ray pair-conversion telescope covering the energy range between
20 MeV and  $\sim$ 1 TeV. Due to its excellent angular resolution ($\sim 0.1^{\circ}$
above 10 GeV), large field of view ($\sim 2.4$ sr), and
very efficient rejection of background from charged particles, it is currently
the best experiment to investigate the nature of the extra-galactic \g-ray background \citep[EGB,][]{2015IGRB} in the GeV energy range.

For our analysis, we have used 78 months of data
from August 4th 2008 to January 31th 2015 ({\it Fermi} Mission Elapsed Time 239557418 s - 444441067 s),
considering the Pass 8 event 
selection\footnote{For a definition of the Pass 8 event selections and their characteristics, see http://www.slac.stanford.edu/exp/glast/groups/canda/ \\ lat\_Performance.htm.}.
Furthermore, to reduce the contamination from the bright Earth limb emission, 
we exclude photons detected
with measured zenith angle larger than 100$^{\circ}$.
In order to generate the final flux maps we have
produced  the corresponding exposure maps using the standard routines from
the LAT \emph{Science Tools}\footnote{http://fermi.gsfc.nasa.gov/ssc/data/analysis/documentation/ \\ Cicerone/} version 10-01-01,
and the Pass 8 CLEAN event class, namely the
{\verb"P8R2_CLEAN_V6"} instrument response functions (IRFs).
We use both back-converting and front-converting events.
The GaRDiAn package \citep{FermiLAT:2012aa,Ackermann:2009zz} was adopted to pixelize both
photon count and exposure maps in HEALPix\footnote{http://healpix.jpl.nasa.gov/}
format \citep{2005ApJ...622..759G}. The maps contain
$N_{\rm pix} = 12, 582, 912 $ pixels with mean spacing  of $\sim 0.06^{\circ}$
corresponding to the HEALPix resolution parameter  $N_{\rm side}=1024$.
Finally,  the flux maps are obtained by dividing the count maps by exposure maps. 
We perform the bulk of our analysis in the three separate  energy
intervals: $0.5<E<1$ GeV, $1<E<10$ GeV and $10<E<100$ GeV.
For a more accurate study of the spectral dependence we will also
use a finer energy binning, with 8 bins {cut at} 0.25, 0.5, 1, 2 , 5, 10, 50, 200, 500 GeV.

Our analysis is focused on the unresolved \g-ray background (UGRB),
i.e., the unresolved EGB emission left after subtracting resolved point sources \citep{2015IGRB}.
{To obtain such maps we mask  out} the \g-ray point sources listed in the 3FGL catalog \citep{3FGL}.
More precisely, we mask the 500 brightest point sources (in terms of the integrated photon flux in the 0.1-100 GeV energy range) 
with a disc of radius $2^\circ$, and the remaining ones with a disc of $1^\circ$ radius.
The Small and Large Magellanic Clouds, which are extended sources, are masked with discs
of  $3^{\circ}$ and $5^{\circ}$  radius respectively.
To reduce the impact of the Galactic emission
we apply a Galactic latitude cut masking the region with \mbox{$|b|< 30^{\circ}$}. 
In \cite{xia11} we have experimented with different latitude cuts and found that \mbox{$|b|>30^{\circ}$}
represents the best compromise between pixels statistics and Galactic contamination.
We also exclude the regions associated
to the {\it Fermi} Bubbles and the Loop~I structures {as in \cite{xia11}}.

The Galactic diffuse emission can be still significant at the high Galactic latitude used in our analysis and 
needs to be removed.
For this purpose, we use the model of Galactic diffuse emission
\verb"gll_iem_v06.fits"\footnote{http://fermi.gsfc.nasa.gov/ssc/data/access/lat/ \\ BackgroundModels.html \label{foot4}},
which we subtract from the observed emission to obtain the {\it cleaned} \g-ray maps.
The model, together with an isotropic template, is convolved with the IRFs and fitted to the photon data in each
energy bin and in our region of interest using  \verb"GaRDiAn". 
The best fit diffuse plus isotropic model is subtracted from the count map and this
residual count map is further divided by the exposure to give the final residual
flux map to be analyzed for the given energy bin.

As the Galactic diffuse emission model is not exact, cleaning is not perfect and the
residual flux maps  are still contaminated by
 spurious signals, especially on large angular scales. 
{However, and this is the main advantage of our analysis,  the cross-correlation analyses are expected to be 
almost immune to these contaminations }
 since Galactic foreground emission
is not expected to correlate with the extragalactic signal that we want to investigate. 
Nonetheless, {to minimize the chance of systematic errors we adopt a conservative approach and, following \cite{xia11} and \cite{Xia:2014}}
 we apply {a further}  cleaning procedure that, using   HEALPix tools, removes all contributions from multipoles up to $\ell=10$.

\begin{figure*}[t]
\vspace{-2cm}
\centering
\includegraphics[width=0.33\textwidth]{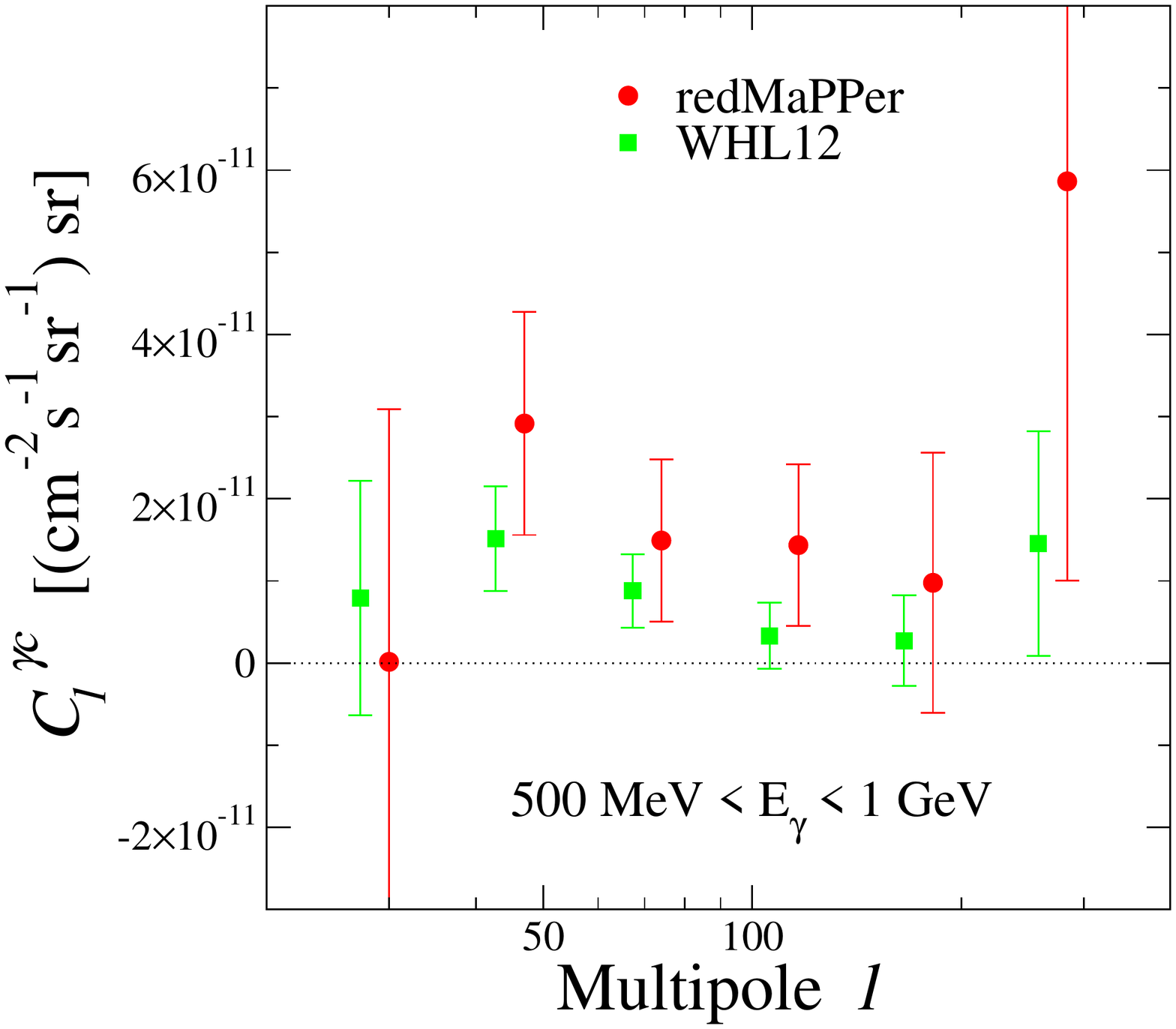}
\includegraphics[width=0.33\textwidth]{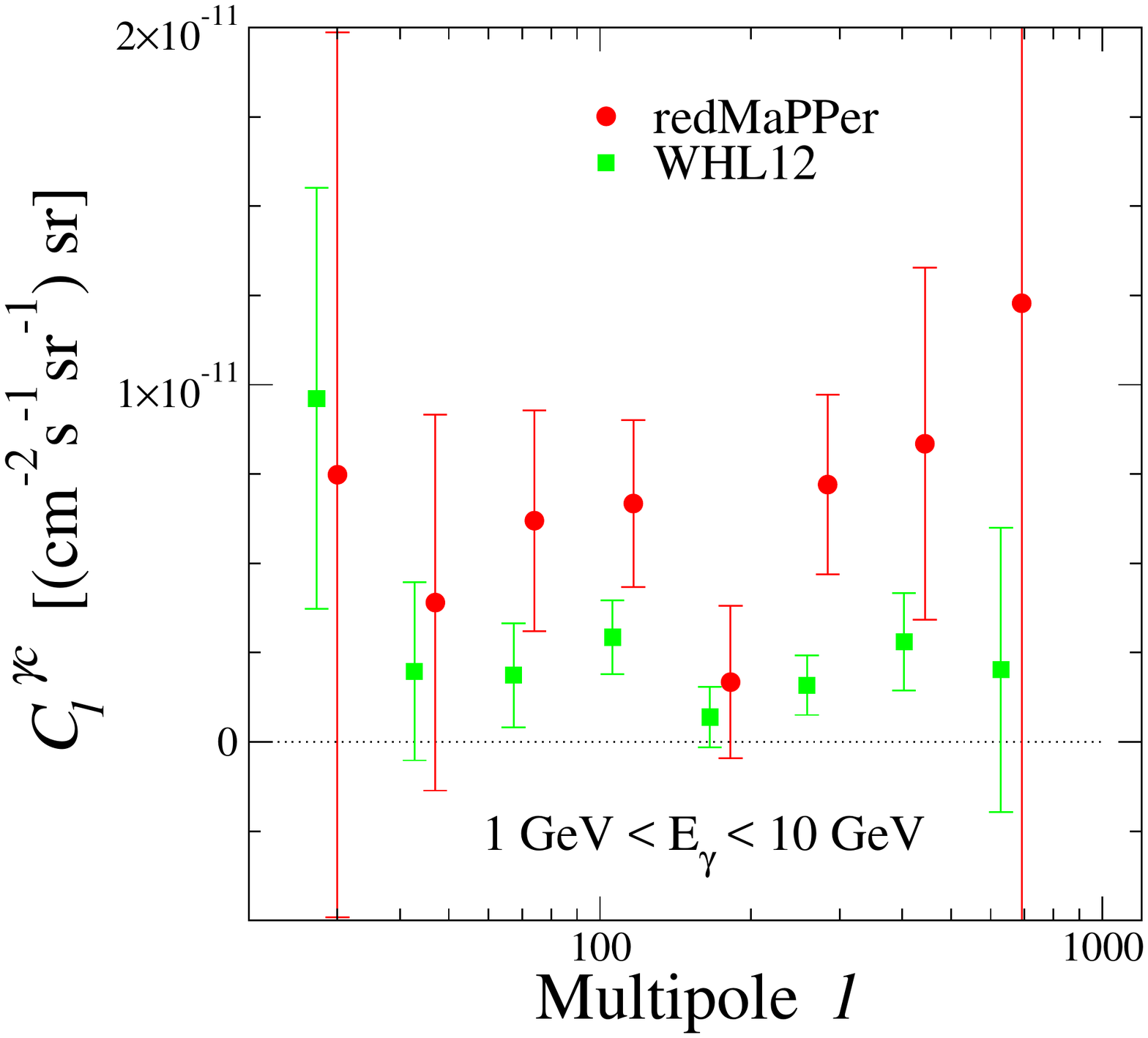}
\includegraphics[width=0.33\textwidth]{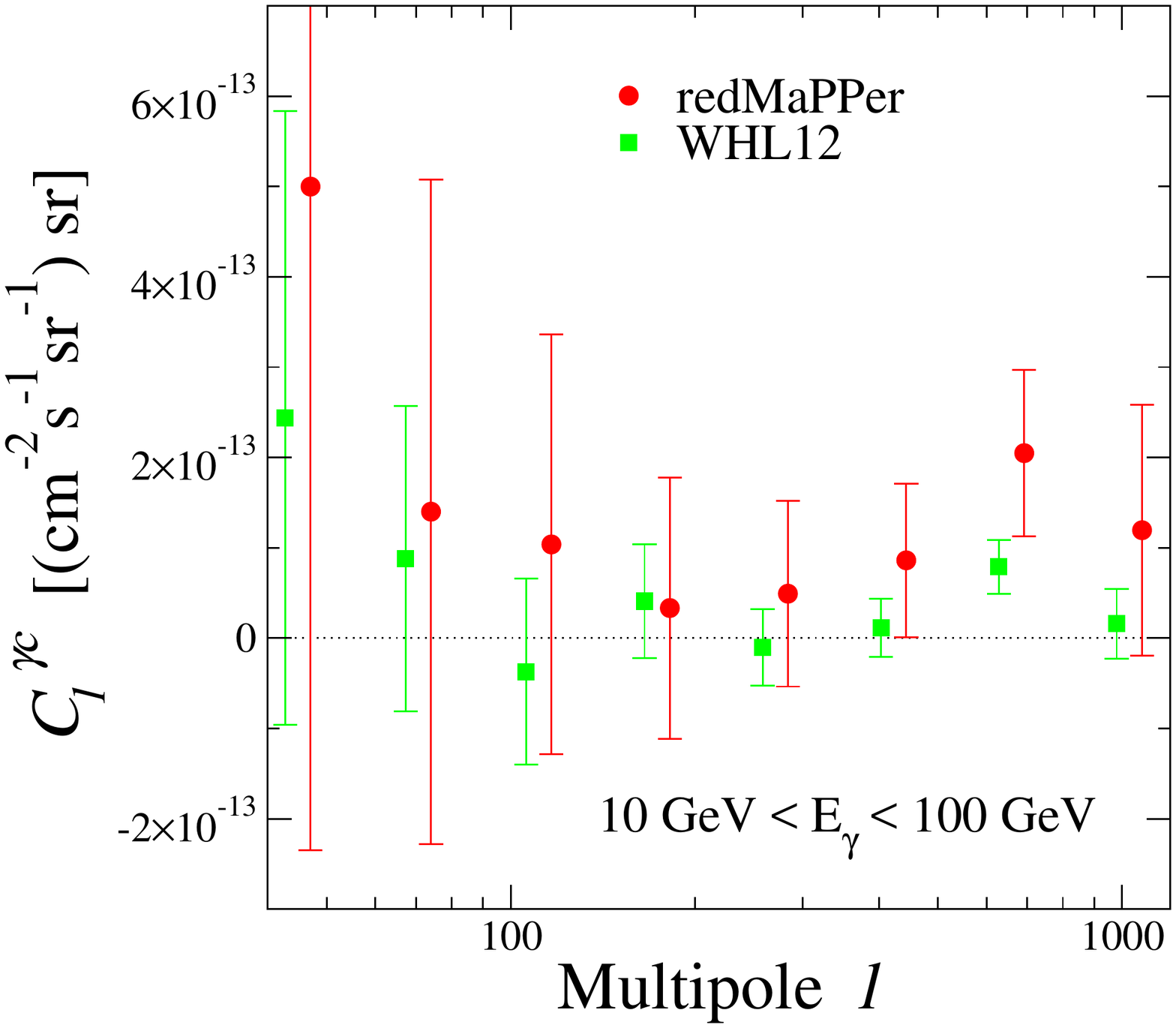}
\caption{Observed CAPS (PSF deconvolved) between the {\it Fermi}-LAT \g-ray map in three different energy bins and the redMaPPer (red) and WHL12 (green) catalogs of clusters. Left panel refers to $500\,{\rm MeV}<E_\gamma<1\,{\rm GeV}$, central panel to $1\,{\rm GeV}<E_\gamma<10\,{\rm GeV}$, and right panel to $10\,{\rm GeV}<E_\gamma<100\,{\rm GeV}$.}
\label{fig:data}
\end{figure*}

\begin{figure}[t]
\vspace{-3cm}
\centering
\includegraphics[width=0.4\textwidth]{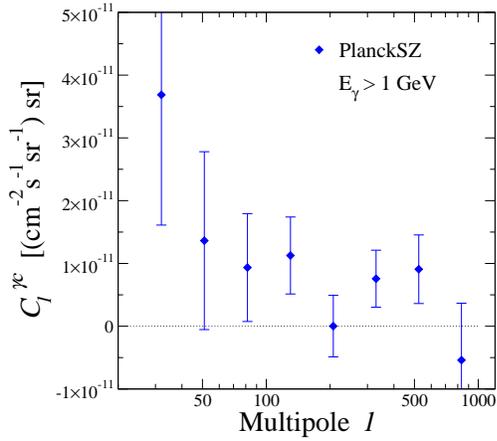}
\caption{Observed CAPS (PSF deconvolved) between the {\it Fermi}-LAT \g-ray map at $E_\gamma>1$ GeV and the PlanckSZ catalog.}
\label{fig:data_Planck}
\end{figure}

\subsection{Galaxy cluster  catalogs}
\label{sec:catadata}

The three catalogs of galaxy clusters considered in our analysis are: 
{\it 1)} SDSS redMaPPer~\citep{2014ApJ...785..104R}, {\it 2)}
WHL12~\citep[][]{2012ApJS..199...34W,2015ApJ...807..178W} and  {\it 3)} PlanckSZ~\citep{2015arXiv150201598P}.

\begin{enumerate}

\item {Clusters in the {\it redMaPPer} catalog are identified using red-sequence galaxies. Adopting these objects as clusters signposts
increases the contrast between cluster and background galaxies in color space, thus enabling accurate and precise photometric redshift estimates. 
In our analysis we consider all  26,350 clusters detected in the redshift range $0.08 < z < 0.55$  over the $\sim$10,400 sq deg area of the SDSS Data Release 8. 
Photo-$z$ errors are nearly Gaussian with an amplitude $\sigma_z \sim 0.006$ at $z \sim 0.1$, that increases to $\sigma_z \sim 0.02$ at $z \sim 0.5$.
Details on the cluster detection procedure and on the iterative method to estimate photometric redshifts can be found in~\citep{2014ApJ...785..104R}.}

\item {{\it WHL12} updates and refines on the \cite{2009ApJS..183..197W}  galaxy cluster catalog from SDSS III~\citep{2011ApJS..193...29A}.} The original catalog contained 39,668 clusters with photometric redshifts. The new catalog was obtained by \cite{2012ApJS..199...34W} by applying 
an improved cluster detection method \citep{2011ApJ...734...68W} to SDSS-III galaxies, exploiting $\sim$1.35 millions  Large Red Galaxies
with spectroscopic redshifts in the SDSS 12th Data Release (DR12). 
The updated WHL12 catalog \cite{2015ApJ...807..178W} that we considered in our analysis has 158,103  clusters in the the range $0.05<z<0.8$.
Its completeness is larger than 95\% for objects with mass $M_{200} > 1.0 \times 10^{14}\, M_\odot$ in the redshift range of $0.05 < z < 0.42$,
decreasing at higher redshifts.

\item  {The {\it PSZ2} Second Planck Sunyaev-Zed'dovich catalog contains SZ-selected clusters. It is 
 based on the full 29 month mission data \citep{2015arXiv150201598P}. 
The methodology employed to detect clusters refines the one used to produce the first Planck SZ cluster catalog. PSZ2's 1,653 clusters, 
distributed across 83.6\% of the sky, are the union of outcomes from three cluster detection codes \citep[see][]{2014A&A...571A..29P}. 
It contains 1653 detections, of which 1203 are confirmed to be clusters with identified counterparts in external datasets and with a purity 
larger than 83\%. PSZ2 probes clusters at relatively low redshift with the distribution peaking at $z\sim0.2$ and extending up to $z\lesssim0.5$. The catalog also provides estimates for the mass of the SZ clusters as a function of redshift.}

\end{enumerate}

For each of the three cluster catalogs we built a HEALPix skymap with resolution parameter $N_{\rm side}=1024$ 
specifying the cluster counts per pixel  $n(\hat{\Omega}_i)$. The three maps are shown in Fig.~\ref{fig:maps}.
Clusters are counted as single objects, i.e. no statistical weight has been used to account for the cluster mass or
selection effects.
The cross-correlation analysis is then performed between the normalized count maps $n(\hat{\Omega}_i)/\bar{n}$, {where $\bar{n}$
is the mean cluster density in the unmasked area}, and the  {\it Fermi}-LAT residual flux sky-maps.
For the two SDSS-based cluster catalogs we use the standard SDSS mask, i.e.,
the contour of the sky region covered by RedMaPPer which can be seen in Fig.~\ref{fig:maps}.
Further, we conservatively mask also the disconnected south-galactic region.
For  the PSZ2 catalog no mask is used since the masked area is included in the one applied to the Fermi maps.

To better understand the catalogs properties, we show in Fig.~\ref{fig:cat_prop} the histograms of the distributions of redshift, mass and angular size of the clusters in the three catalogs.
We note already here that the vast majority of the considered clusters are effectively point-like for the {\it Fermi}-LAT telescope, namely, their angular size is smaller than the  point-spread function (PSF) of {\it Fermi}-LAT (see right panel).

\begin{figure*}[t]
\vspace{-2cm}
\centering
\includegraphics[width=0.33\textwidth]{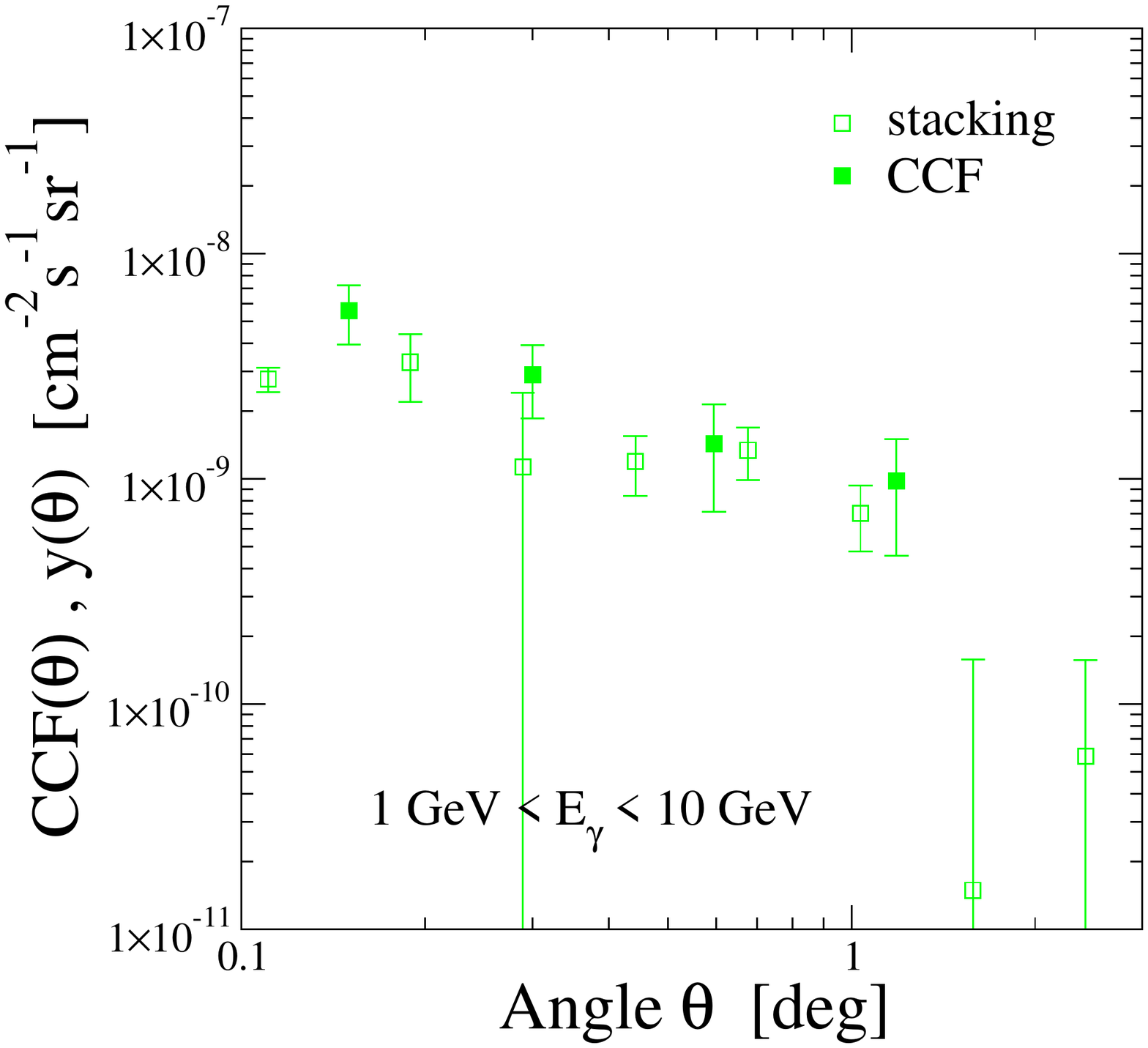}
\includegraphics[width=0.33\textwidth]{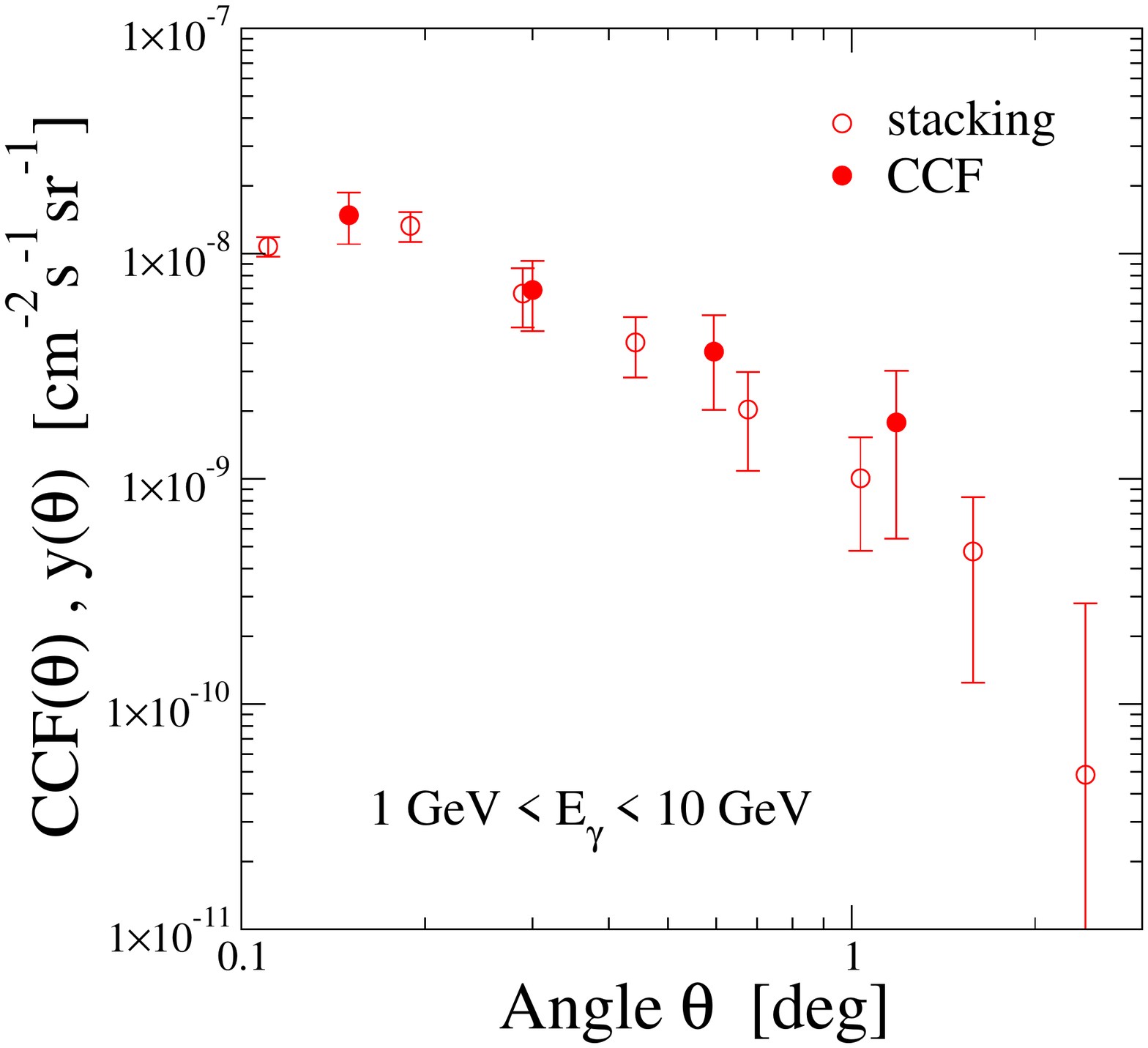}
\includegraphics[width=0.33\textwidth]{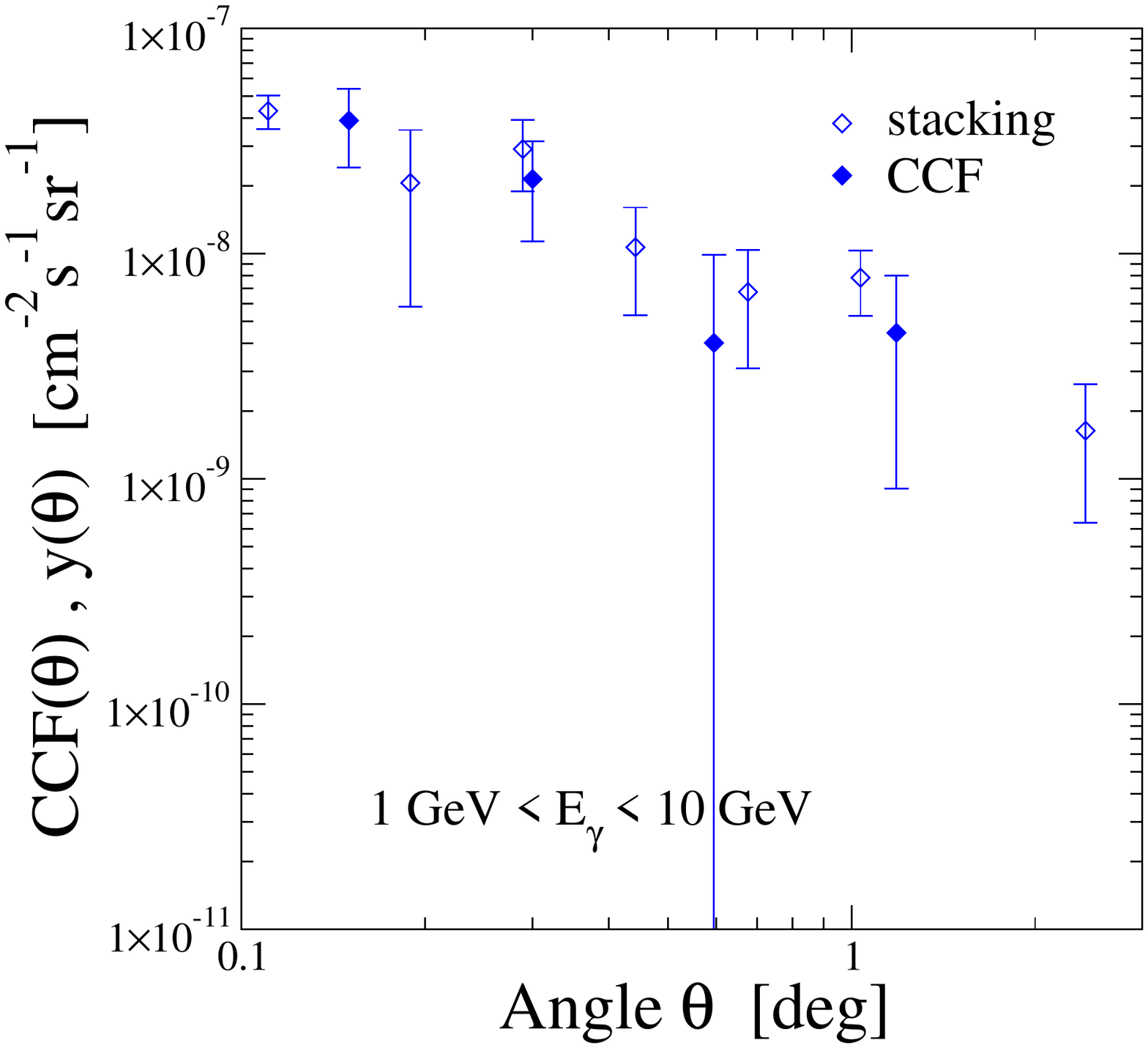}
\caption{Observed CCF (filled points, without PSF deconvolution) between the {\it Fermi}-LAT \g-ray map at  $1 < E_\gamma < 10$ GeV and catalogs of clusters, as a function of the angular separation in the sky, compared to the observed cluster \g-ray stacked profiles (open points), as a function of the angle from the center of the stacking. The WHL12 case is shown in the left panel, redMaPPer in the central panel, and PlanckSZ in the right panel.
Even though the stacked profiles agree well with the CCF, they are not used for quantitative analyses since the profile and
its error bars rely on the assumption that all stacked fields are independent, while in reality they are likely not (see text for details).
}
\label{fig:CCF_stack}
\end{figure*}

\section{Measuring 2-point cross-correlation statistics.}
\label{sec:meas}

\begin{figure*}[t]
\centering
\includegraphics[width=0.3\textwidth]{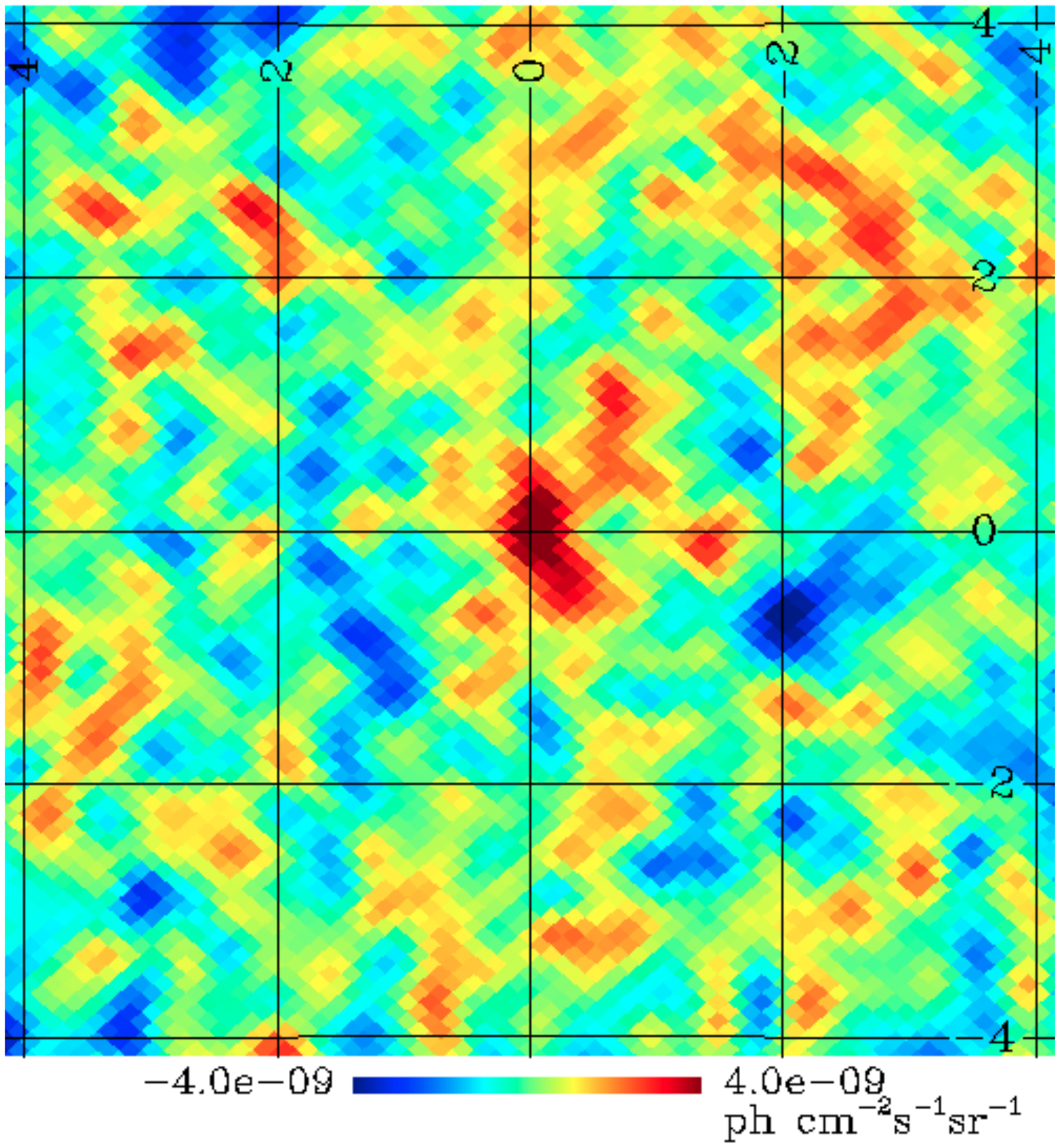}
\hspace{3mm}
\includegraphics[width=0.3\textwidth]{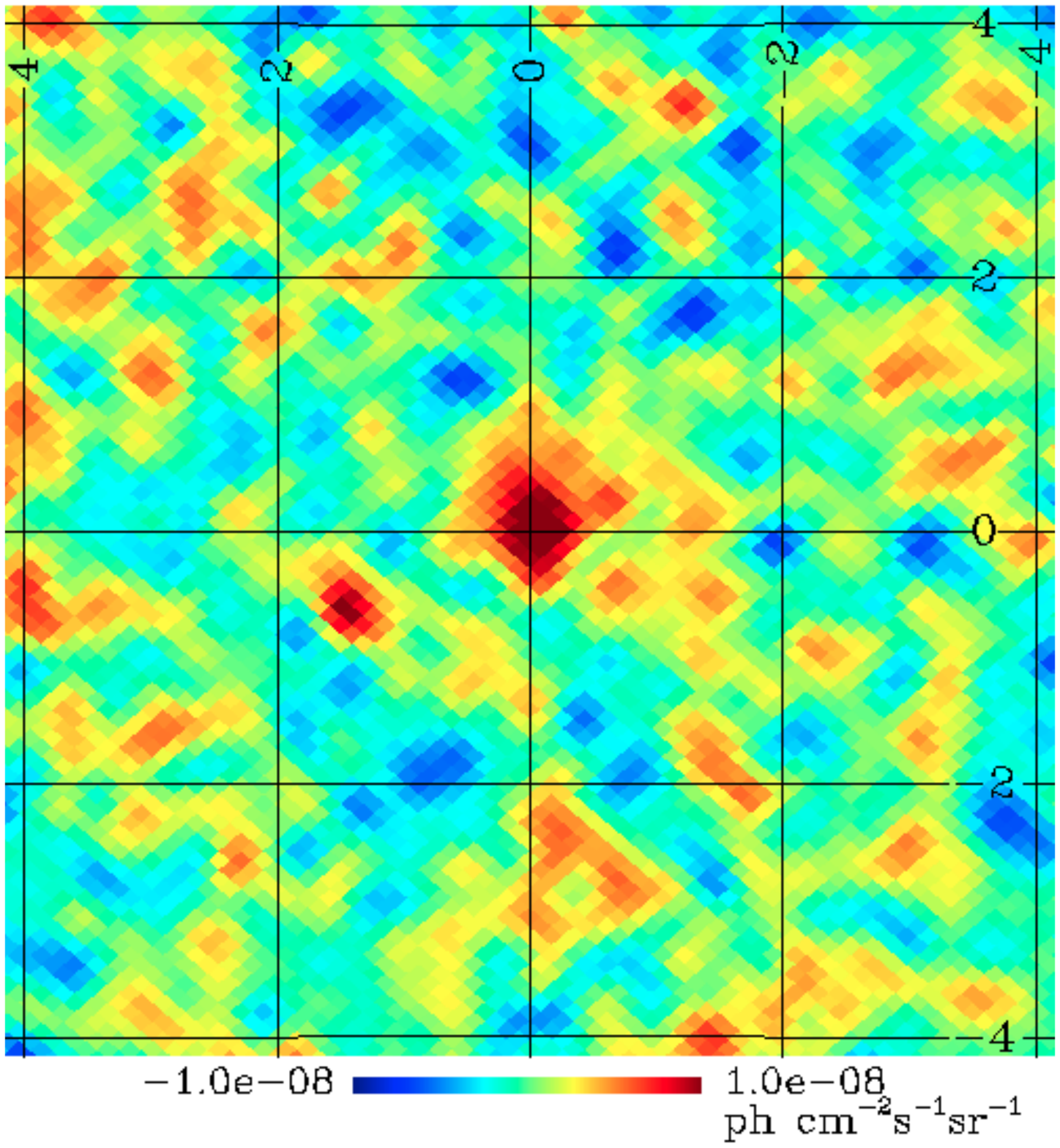}
\hspace{3mm}
\includegraphics[width=0.3\textwidth]{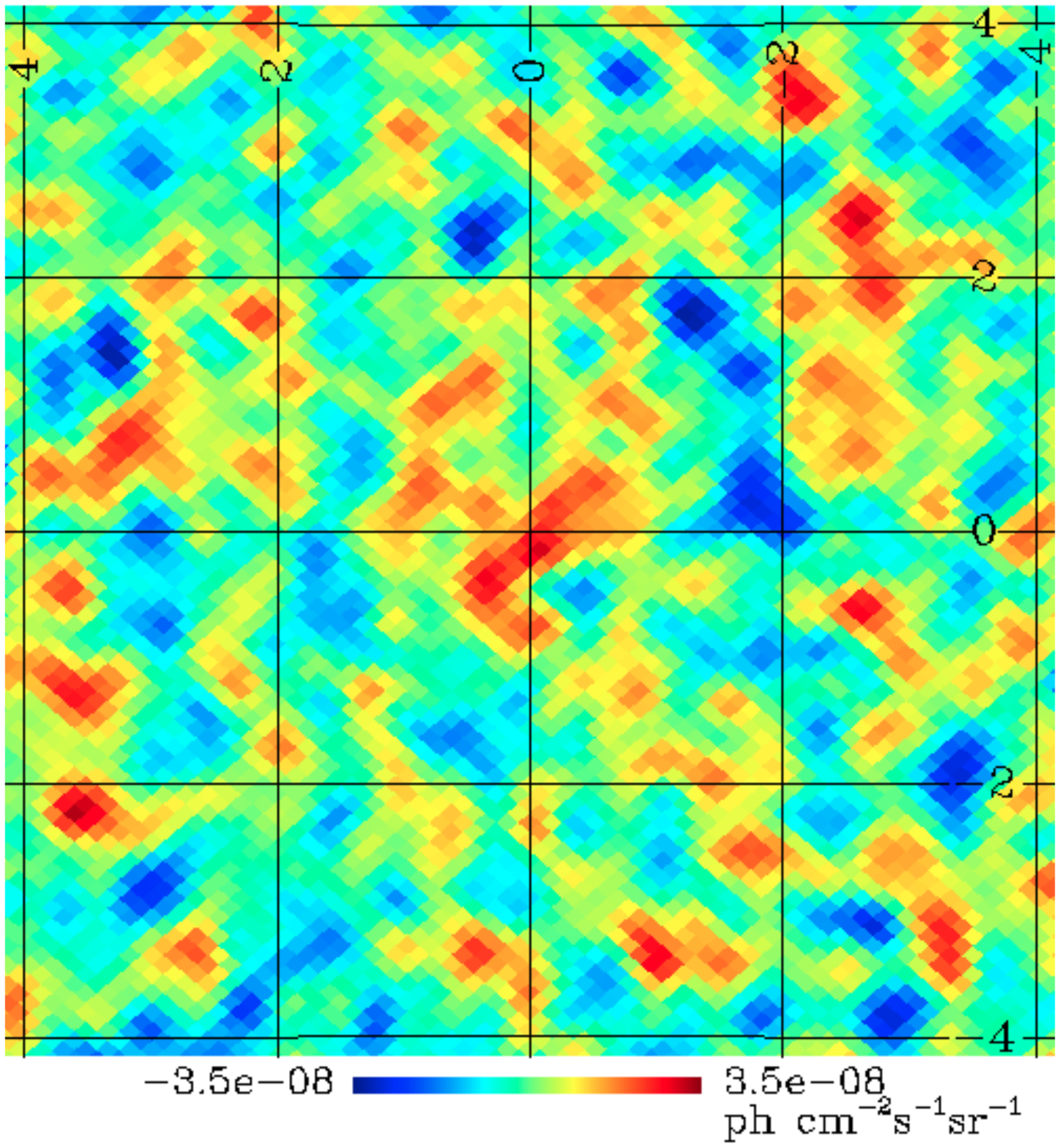}
\caption{Observed cluster \g-ray stacked images for $1 < E_\gamma < 10$ GeV for the catalogs WHL12 (left), redMaPPer (central) and PlanckSZ (right).
}
\label{fig:stack_im}
\end{figure*}

\begin{figure}[t]
\vspace{-3cm}
\centering
\includegraphics[width=0.49\textwidth]{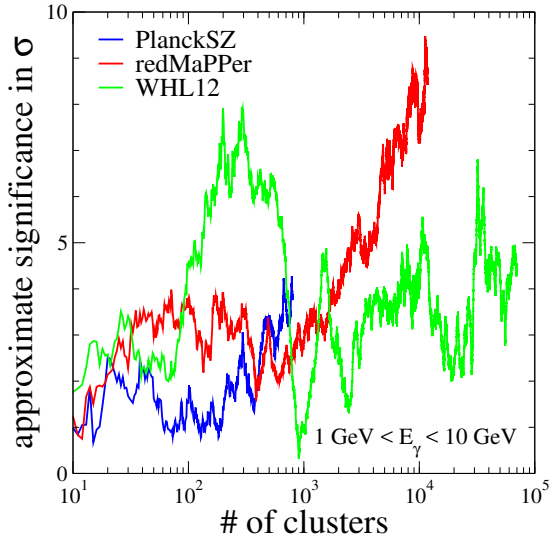}
\caption{Approximate significance of the stacking signal (see text for the definition) as a function of the number of clusters in the stacking for the WHL12 (green), redMaPPer (red) and PlanckSZ (blue) catalogs.
This plot is for illustrative purposes only and is not used for any quantitative analysis.
}
\label{fig:stack_sig}
\end{figure}

\subsection{Cross-correlation APS}
\label{sec:xcorr}

{In our analysis we estimate both the 2-point CCF and the cross APS. In both cases we 
used  PolSpice,\footnote{http://www2.iap.fr/users/hivon/software/PolSpice/},} a publicly available
toolkit to estimate the angular
$CCF^{(\gamma c)}(\theta)$  and the CAPS $\bar C_\ell^{(\gamma c)}$  
of any two datasets pixelized in HEALPix format \citep{szapudi01,chon04,efstathiou04,challinor05}.
{\cite{Xia:2014} have tested the  reliability and  robustness of the PolSpice estimator in similar analyses.}
PolSpice also estimates the covariance matrix $\bar V_{\ell\ell'}$ of the different multipoles 
taking into account the correlation effect induced by the mask.

In general, other, and possibly more rigorous, statistical techniques (see, e.g., \citep{Baddeley,Pewsey}) can be applied. However, for the specific problem under investigation (affected by large uncertainties), they are expected not to affect significantly the results as they could do in the case of a precision test.

The CAPS estimated from PolSpice {include the effects of the instrument  PSF
and pixelization}.  
We  deconvolve the results from these effects as in \cite{Xia:2014}:
Firstly  we compute the beam window function $W_\ell^{B}$ associated to the PSF, and
the pixel window function $W_\ell^{\rm pixel}$ associated to the pixelization.
Then, we derive the deconvolved CAPS $C_\ell^{(\gamma c)}$ from the measured ones $\bar C_\ell^{(\gamma c)}$ as 
$C_\ell^{(\gamma c)}=(W_\ell)^{-1}\,\bar C_\ell^{(\gamma c)}$, where $W_\ell = W_\ell^{B} W_\ell^{\rm pixel} $
is the global window function.
The covariance matrix of the deconvolved  $C_\ell^{(\gamma c)}$ is then given by  
$V_{\ell\ell'}  = \bar V_{\ell\ell'} W_\ell^{-2}W_{\ell'}^{-2}$.
Finally, since the mask induces a strong correlation in nearby multipoles we bin the CAPS measurements
into 12 equally spaced logarithmic intervals in the range $\ell\in[10,2000]$. 
We choose logarithmic bins to account for the rapid loss of power at high $\ell$ induced by the PSF.
In what follows we omit the square bracket in the $[\ell]$ subscript and use $C_\ell^{(\gamma c)}$ also to indicate the binned CAPS.
In our analysis we mainly focus on the binned quantity and should be clear from the context when, instead, we 
consider the un-binned CAPS.
The $C_\ell^{(\gamma c)}$ in each bin is given by the simple unweighted average
of the $C_\ell^{(\gamma c)}$ within the bin.
For these binned $C_\ell^{(\gamma c)}$ it is also possible to build
the corresponding block covariance matrix  as $\sum_{\ell\ell'} V_{\ell\ell'}/\Delta\ell/\Delta\ell'$,
where $\Delta\ell, \Delta\ell'$ are the width of the two multipoles bins, and $\ell, \ell'$ run within
the multipoles of the first and second multipole bin.

We verified that the binning is very efficient in removing correlation among nearby multipoles, 
resulting in a block covariance matrix that is, to a good approximation, diagonal. For this reason
we have neglected the off-diagonal terms in our analysis.

The  results of the various CAPS measurements are shown in Figs.~\ref{fig:data} and \ref{fig:data_Planck}.
In Fig.~\ref{fig:data} we show the PSF-deconvolved and binned CAPS of the redMaPPer (red) and WHL12 (green) catalogs
in threee energy bins: $500\,{\rm MeV}<E_\gamma<1\,{\rm GeV}$ (left panel), 
$1\,{\rm GeV}<E_\gamma<10\,{\rm GeV}$ (central) and $10\,{\rm GeV}<E_\gamma<100\,{\rm GeV}$ (right).
The bars represent 1 $\sigma$ errors from the diagonal elements of the covariance matrix.

We clearly detect a non-zero cross-correlation signal. We quantify the local statistical significance of the detection in each energy bin 
in terms of number of sigmas  $N_\sigma=\sqrt{\sum_j (C_{\Delta\ell_j}^{(\gamma c)}/\delta C_{\Delta\ell_j}^{(\gamma c)})^2}$ 
with $C_{\Delta\ell_j}^{(\gamma c)}$ the measured CAPS in the $j$ bin  
and $\delta C_{\Delta\ell_j}^{(\gamma c)}$ its uncertainty quoted in the figure.
For WHL12 we find: $N_\sigma=3.7, \, 4.4$ and 2.9 in the three energy bins. 
The significance for  redMaPPer is very similar:  $N_\sigma=3.3, \, 5.0$ and 2.7.

The results for PlanckSZ are shown in  Fig. \ref{fig:data_Planck}. Given the limited number of 
clusters in this sample the statistics is too poor to divide results into energy bins. Therefore, we have 
considered a single bin $E_\gamma>1$ GeV. Also in this case we find a non-negligible cross correlation signal with
$N_\sigma=3.7$.

We note that the amplitude of the measured CAPS is the highest for PlanckSZ and the lowest for WHL12, with redMaPPer being in between.
This was somehow expected since the average mass of the clusters in the catalog is the largest in PlanckSZ and the smallest in WHL12 (see Fig.~\ref{fig:cat_prop}b).
Since the redshift distribution of the clusters is not dramatically different, larger mass (typically) implies higher \g-ray emission and higher catalog bias, and therefore higher CAPS amplitude.

\subsection{CCF and Stacked signals}
\label{sec:stack}
{The CCF statistics provides a complementary information to CAPS.
Here we estimate CCF from the CAPS as follows:}
\be
 CCF^{(\gamma c)}(\theta) = \sum_\ell\frac{2\ell+1}{4\pi}\bar C_\ell^{(\gamma c)} P_\ell[\cos(\theta)] \;,
\label{eq:2point}
\ee
where $\theta$ is the angular separation in the sky and $P_\ell$ are the Legendre polynomials.

The resulting CCFs from the three cluster catalogs are shown Fig.~\ref{fig:CCF_stack}. 
{We show the 
$1 < E_\gamma < 10$ GeV case only. Unlike the CAPS case, here the CCFs {\it are not de-convolved} for the PSF and 
angular pixels. The reason for this is that unlike the Fourier space case, in which 
deconvolution is a simple multiplication,  deconvolution in configuration space is more unstable if, like 
in our case,  CAPS has large power at high multipoles $\ell$.}

The CCF analysis is quite similar to stacking the  \g-ray signal  at clusters' locations (as we will show at the end of the Section).
Since stacking analyses have been popular in previous studies we 
decided to also perform this type of analysis.
In our procedure we first select a region of {4 degrees} of radius around the position of each cluster in the three catalogs.
Then, we sum the \g-ray flux in the circular areas with no attempt to rescale the signal to the object's properties (e.g. richness).  Each image in the stacking is randomly rotated, in order to better investigate the impact of the region around the centered cluster. A potential  issue is that the stacked images might not be independent, since they come from partially
overlapping fields. This is most severe in the WHL12 catalog in which  the mean 
angular separation of clusters is $\sim 0.25^\circ \ll 4 ^\circ$ radius size. 
For this reason we will not attempt to perform a quantitative statistical comparison with the results of the CCF analysis.

The stacked images for \g-rays in the energy range $1 < E_\gamma < 10$ GeV are shown in Fig.~\ref{fig:stack_im}. We see a clear \g-ray excess at the clusters' center.
From each image we have obtained a stacked  \g-ray emission profile $y^{(\gamma c)}(\theta)$
by averaging the stacked flux in logarithmically-spaced circular annuli ignoring possible small anisotropies that may survive the stacking 
procedure.
The \g-ray stacked profiles for the clusters in the three catalogs are shown in Fig.~\ref{fig:CCF_stack}.
The reported error bars are the image-by-image scatter 
around the stacked flux in each annulus.
With the same definition of errors, we show how the stacking signal builds up as the number of clusters increases in Fig.~\ref{fig:stack_sig}. We defined an approximate significance given by $\sqrt{\sum_i\,(y_i/\sigma_i)^2}$, where $y_i$ is the measured stacked emission in the annulus $i$ and $\sigma_i$ is the scatter.
On the other hand, these approximations are likely to underestimate the real errors and to overestimate the significance, since rely on the assumption that all stacked fields are independent, while in reality they are not, as already mentioned.
Therefore both Fig.~\ref{fig:CCF_stack} and the insets of Fig.~\ref{fig:stack_sig} are shown only to illustrate the trends, and will not be used for any quantitative analyses.

If the \g-ray emission is circularly symmetric, the stacked profile $y^{(\gamma c)}(\theta)$ (derived with the procedure outlined above) is an estimator of $\langle \delta_c(0)\,I_\gamma(\theta)\rangle$, where $\delta_c$ is the cluster fluctuation field. This is actually the definition of the CCF, i.e. $CCF^{(\gamma c)}(\bm \theta)= \langle \delta_c(\bm \theta')\,I_\gamma(\bm \theta'+\bm \theta)\rangle=\langle \delta_c(0)\,I_\gamma(\theta)\rangle$.
Therefore, in the small angle limit, where one can assume the dominant contribution to come from a single object and the circular symmetry to hold, the two quantity $CCF^{(\gamma c)}$ and $y^{(\gamma c)}(\theta)$ are perfectly equivalent. 
\footnote{In the literature, the relation between the stacked profile and the CAPS is typically reported using the Fourier transform (see, e.g., \cite{2012PhRvD..85b3007F}):
\be
 y^{(\gamma c)}(\theta) = \int \frac{d\ell\,\ell}{2\pi}\bar C_\ell^{(\gamma c)} J_0(\ell \theta) \;,
\label{eq:ystack}
\ee
where $J_0$ is the is the zeroth order Bessel function.
In the small angle limit $\ell \gg1$, $\theta \ll1$ we have $J_0(\ell \theta)\simeq P_\ell[\cos(\theta)]$ and thus Eq.~(\ref{eq:2point}) becomes the discrete form of Eq.~(\ref{eq:ystack}).}

The similarity between the two quantities is 
indeed evident from Fig.~\ref{fig:CCF_stack}.
The nice agreement between CCFs and stacking profiles is an important cross-check for our analysis, since the two measurements have been obtained employing two completely different methods.

\section{Models}
\label{sec:models}

In the Limber approximation \citep{1953ApJ...117..134L}, the CAPS between \g-ray emitters and galaxy clusters can be written as:
\be
 C_\ell^{(\gamma c)}=\int \frac{d\chi}{\chi^2} W_{\gamma}(\chi)\, W_{c}(\chi)\,P_{\gamma c}\left(k=\ell/\chi,\chi\right)\;,
\label{eq:clfin}
\ee
where $\chi(z)$ denotes the radial comoving distance.
$W_c(\chi)$  and $W_{\gamma}(\chi)$ are the window functions that characterize the distribution of clusters  and \g-ray emitters
along the  line of sight, respectively.
$P_{\gamma c}(k,z)$ is the 3D cross power spectrum (PS) at the redshift $z$,  $k$ is the modulus of the wavenumber and $\ell$ denotes the angular multipole.  
The relation $\chi(z)$ is fully specified by the expansion history of the Universe $H(z)$:
$d\chi=c\,dz/H(z)$.

In the following, galactic \g-ray astrophysical emitters are denoted by the symbol $\gamma_i$, where $i$ indicates the type of source, and we consider the ones that 
are known to significantly contribute to the UGRB signal: blazars, misaligned AGN [mAGN] and star forming galaxies [SFGs].
The symbol $\gamma_c$ refers to \g-ray emission from the intra-cluster medium [ICM], while $c_j$ denotes cluster catalogs ($j=$ redMaPPer, WHL12 or PlanckSZ).

The ingredients of our model are the cross-PS and the window functions entering in the computation of Eq.~(\ref{eq:clfin}). They are described in the next two subsections.
Predictions of Eq.~(\ref{eq:clfin}) will be then compared with the measured  CAPS shown in Figs.~\ref{fig:data} and \ref{fig:data_Planck}.

\subsection{Window functions}
\label{sec:wf}
The window functions of the three galactic \g-ray sources considered here (blazars, mAGNs and SFGs)
are presented in \cite{2015ApJS..221...29C} (Appendix).
The collective \g-ray emission from ICM in the Universe gives raise to the window function:
\be
W_{\gamma_c}(z)=\int_{M_{c,min}} \,\de M\,\frac{d^2n}{dM\,dV}\,\frac{\mathcal{L}_{\gamma_c}(M_{500}(M),z)}{4\pi\,(1+z)}\;,
\label{eq:wf_clug}
\ee
where $d^2n/dM\,dV$ is the cluster mass function predicted by the ellipsoidal collapse model \citep{Sheth:1999mn}
and $\mathcal{L}_{\gamma_c}$ is the cluster \g-ray luminosity per unit energy range.
The integral is over the cluster masses above $10^{13.8}\,h^{-1}\,M_\odot$ \citep{2015A&A...578A..32Z}.
The relation between the halo and the cluster virial mass $M_{500}$ is specified in \cite{2003ApJ...584..702H} (Appendix A).
For the luminosity we adopt the empirical relation of \cite{2015A&A...578A..32Z}: 
\be
L_{\gamma_c}(100\,{\rm MeV})=A_{\gamma_c}\,\left(\frac{M_{500}}{M_\odot}\right)^{5/3}\,{\rm s^{-1}\,GeV^{-1}}\;,
\label{eq:lumg_clu}
\ee
where  $L_{\gamma_c}=\mathcal{L}_{\gamma_c}/E$ 
and  $A_{\gamma_c} \lesssim10^{21}$ is a normalization parameter constrained by the non-detection of  
\g-rays from nearby clusters ~\citep{2015A&A...578A..32Z}.
Finally, we assume a power-law ICM energy spectrum with index $\alpha_{\gamma_c}=2.2$.

The window function of cluster counts is:
\be
W_{c_j}(z)=\frac{4\pi\,\chi(z)^2}{N_{c_j}}\int \,\de M\,\frac{d^2n_{c_j}}{dM\,dV}\;,
\label{eq:wf_cluc}
\ee
where:
\be
N_{c_j}=\int \,\de M\,dV\,\frac{d^2n_{c_j}}{dM\,dV}
\ee
is the number of object in the $j$-th cluster catalog.
To model the cluster mass function we adopted two different approaches.

{
For the redMaPPer and WHL12 clusters
we measure the cluster richness distribution $d^2n_{c_j}/{d\lambda dz}$ 
from the catalogs, assume a lognormal
distribution for the cluster richness $\lambda$, $P(\lambda|M_{500})$, with $\lambda(M_{500})$ taken from 
\cite{2015MNRAS.454.2305S} (redMaPPer) and  \cite{2015ApJ...807..178W} (WHL12) and derive the cluster mass function 
as a function of redshift as:
\be
\frac{d^2n_{c_j}}{dz\,dM_{500}}=\int d\lambda \frac{d^2n_{c_j}}{d\lambda dz} P(\lambda|M_{500}) \, .
\ee
Finally, we obtain the mass function per unit volume $d^2n_{c_j}/{dV dM}$
by specifying the cosmology-dependent relation:
\be
\frac{d^2V}{d\Omega dz}=\frac{\chi(z)^2}{H(z)}\, .
\ee
We checked that the derived cluster mass function is in good agreement with the theoretical model of \cite{Sheth:1999mn}, in the relevant mass 
and redshift ranges, once selection effects and completeness are taken into account.

For the PlanckSZ clusters we estimate the mass function using directly the masses and the redshifts of the objects in the catalog.}

\begin{figure}[t]
\vspace{-3cm}
\centering
\includegraphics[width=0.49\textwidth]{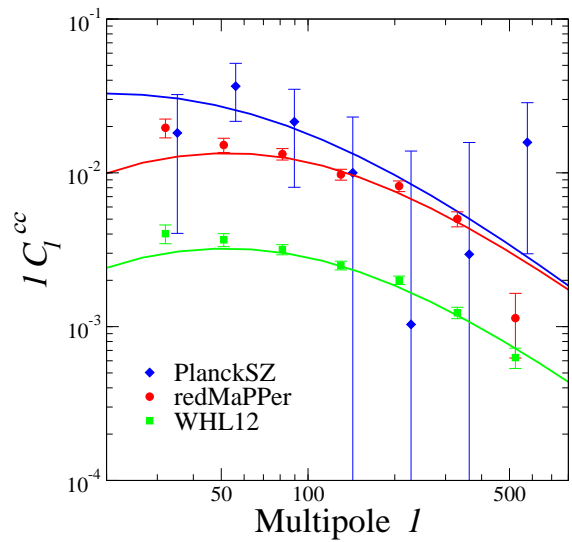}
\caption{Measured (points) and predicted (lines) autocorrelation APS for the catalogs redMaPPer (red), WHL12 (green) and PlanckSZ (blue).}
\label{fig:APS_catalogs}
\end{figure}

\subsection{Three dimensional power spectrum}
\label{sec:ps}

To model the three dimensional cross power spectrum we use the halo model and write the PS as a sum of 
the one-halo and two-halo terms $P=P^{1h}+P^{2h}$. 
Below we provide the expressions for $P^{1h}$ and $P^{2h}$. The interested reader can find a detailed discussion in \cite{Fornengo:2013rga}.
We also notice that in some equation the redshift dependence is not explicitly reported.

For the cross correlation between point-like astrophysical \g-ray emitters and clusters we have:
\bea
& & P_{c_j,\gamma_i}^{1h}(k,z) = \int_{\mathcal{L}_{\rm min,i}(z)}^{\mathcal{L}_{\rm max,i}(z)} \de \mathcal{L}\,\Phi_i(\mathcal{L},z)\,\times \nonumber\\
& & \qquad \qquad \qquad  \frac{\mathcal{L}}{\langle f_{\gamma_i} \rangle} \,\frac{\langle N_{c_j}\!(M(\mathcal{L}))\,\rangle}{\bar n_{c_j}} \label{eq:PSBd1} \\
& & P_{c_j,\gamma_i}^{2h}(k,z) = \left[\int_{\mathcal{L}_{\rm min,i}(z)}^{\mathcal{L}_{\rm max,i}(z)} \de \mathcal{L}\,\Phi_i(\mathcal{L},z)\, b_{\gamma_i}(\mathcal{L})\,\frac{\mathcal{L}}{\langle f_{\gamma_i} \rangle} \right]\nonumber\times\\
& &\qquad \left[\int_{M_{\rm min}}^{M_{\rm max}} \de M\,\frac{dn}{dM} b_h(M)\,\frac{\langle N_{c_j}\,\rangle}{\bar n_{c_j}} \right] \,P^{\rm lin}(k)\;.\label{eq:PSBd2}
\eea
Both terms depend on the luminosity function of the emitter, $\Phi_i$ and the mean luminosity density
 $\langle f_{\gamma_i}\rangle = \int \de \mathcal{L} \, \mathcal{L} \, \Phi_i(\mathcal{L},z)$, 
whereas the linear mass power spectrum $P^{\rm lin}(k)$
and the bias $b_{\gamma_i}$ only enter the two-halo term. For the bias we adopt a simple linear model and assume that the 
bias of the emitter is equal to that of its halo host $b_{\gamma_i}(\mathcal{L})=b_h(M(\mathcal{L}))$ modeled according to 
\cite{Sheth:1999mn}. 
For the relation between the mass of the halo host and the luminosity of the emitter, 
$M(\mathcal{L})$, we adopt the one derived by \cite{Camera:2014rja}.
The effective halo occupation of clusters $\langle N_{c_j}\rangle=(dn_{c_j}/dM)/(dn/dM)$ is obtained from the cluster mass functions 
$dn_{c_j}/dM$  used in Section~\ref{sec:wf}. In this way, we account for selection effects and completeness of the catalogs.
The average number density of clusters at a given redshift is given by $\bar n_{c_j}(z)=\int dM\, \langle N_{c_j}\rangle\,dn/dM$.
Note that Eq.~(\ref{eq:PSBd1}) does not depend on the wavenumber $k$. It describes the picture of point-like \g-ray emitters located at the center of the clusters. Being flat, it acts as a shot-noise-like term.

{\cite{2015ApJS..221...29C} have shown that this halo model is not sufficient to describe the effect of the {\it Fermi}-LAT PSF 
that creates an additional shot-noise-like term on small-scales, which is not captured by the above equations. 
Quantifying the amplitude  
of this effect is not straightforward. However, since we know it is scale-independent, we can model it empirically by adding an extra, shot-noise-like
constant term in the fit of the measured $C_{\ell}^{(\gamma c)}$. Therefore, following \cite{2014JCAP...10..061A} and \cite{2015ApJS..221...29C}
we include one additional free parameter  for each combination of cluster catalog and \g-ray source.}

We note also that the 1-halo term model above assumes that the relation $M(\mathcal{L})$ is deterministic. We argue that ignoring the scatter in the relation
does not significantly affect our results since the 1-halo term is small (blazars, mAGN and SFG reside in halos typically smaller than the cluster size)
and subdominant with respect to the shot-noise term.

For the cross correlation between \g-ray emission from the ICM and clusters we have:
\bea
& & P_{c_j,\gamma_c}^{1h}(k,z) = \int_{M_{\rm c,min}}^{M_{\rm max}} \de M\ \frac{dn}{dM} \frac{\langle N_{c_j}\,\rangle}{\bar n_{c_j}} \frac{\mathcal{L}_{\gamma_c}(M)}{\langle f_{\gamma_c} \rangle}\, \frac{\tilde v_\delta(k|M)}{M} \nonumber \\ 
& & P_{c_j,\gamma_c}^{2h}(k,z) = \left[\int_{M_{\rm min}}^{M_{\rm max}} \de M\,\frac{dn}{dM} b_h(M) \frac{\langle N_{c_j}\rangle}{\bar n_{c_j}} \right]\times \label{eq:PSclcl2}\\
& & \left[\int_{M_{\rm c,min}}^{M_{\rm max}} \de M \,\frac{dn}{dM} b_h(M)\,\frac{\mathcal{L}_{\gamma_c}(M)}{\langle f_{\gamma_c} \rangle}\,\frac{\tilde v_\delta(k|M)}{M} \right]\,P^{\rm lin}(k) \;,\nonumber
\eea
where now the luminosity density is $\langle f_{\gamma_c}\rangle = \int \de M \,dn/dM\, \mathcal{L}/\bar\rho$, and $\tilde v_\delta(k|M)$ is the Fourier transform of
the normalized halo density profile  $\rho_h(\bm x|M)/\bar \rho_{DM}$, that we assume to have a NFW shape~\citep{Navarro:1996gj}.
The underlying assumption is that  the \g-ray emission from the ICM has the same profile of the host halo (in practice, this is not a crucial assumption, since in the current analysis we do not probe scales smaller than the typical size of a cluster).

Unlike in the previous case, uncertainties in the 1-halo term cannot be ignored. They stem from the fact that no extended \g-ray emission from clusters has been unambiguously detected and, consequently,  no observational constraint exists for the $\mathcal{L}_{\gamma_c}(M)$ relation.
To account for this potential source of systematic error, we again include an additional constant term when we fit the cross-correlation model to the data.

\begin{figure}[t]
\vspace{-3cm}
\centering
\includegraphics[width=0.49\textwidth]{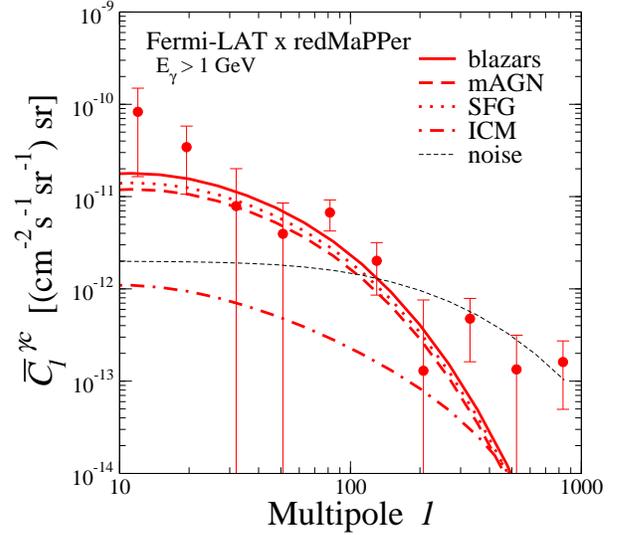}
\caption{CAPS between redMaPPer clusters and \g-ray emitters (convolved with beam window function $W_\ell^{B}$) compared to the measurement at $E_\gamma>1$ GeV. Dotted line shows a noise term at the level of $C_\ell^{(\gamma c)}=2\cdot10^{-12}{\rm cm^{-2}s^{-1}}$.}
\label{fig:CAPS_redMA}
\end{figure}

\begin{figure}[t]
\vspace{-3cm}
\centering
\includegraphics[width=0.49\textwidth]{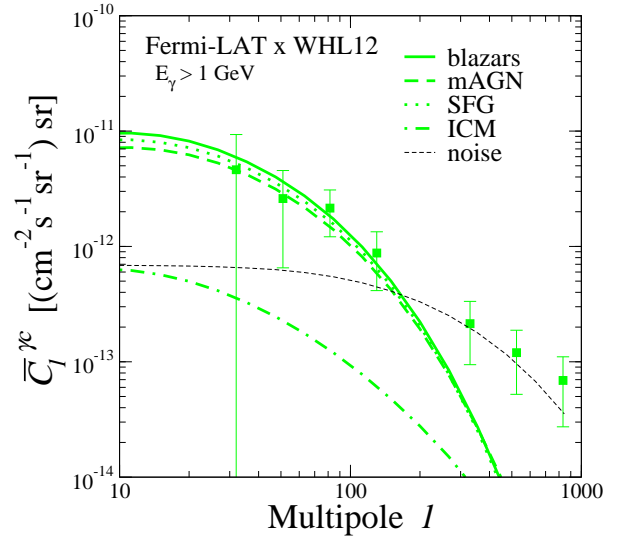}
\caption{CAPS between WHL12 clusters and \g-ray emitters (convolved with beam window function $W_\ell^{B}$) compared to the measurement at $E_\gamma>1$ GeV. Dotted line shows a noise term at the level of $C_\ell^{(\gamma c)}=7\cdot10^{-13}{\rm cm^{-2}s^{-1}}$.}
\label{fig:CAPS_WHL12}
\end{figure}

\section{Results}
\label{sec:res}

Before comparing model and data we performed a sanity check in which we use our model to predict the angular cluster-cluster power spectra and compare it with 
the measured auto angular power spectrum for each of the three cluster catalogs. The results are shown in   Fig.~\ref{fig:APS_catalogs}.
The agreement is remarkably good except at very large or very small angular scales where, respectively, selection effects and model uncertainties are larger.

\begin{figure}[t]
\vspace{-3cm}
\centering
\includegraphics[width=0.49\textwidth]{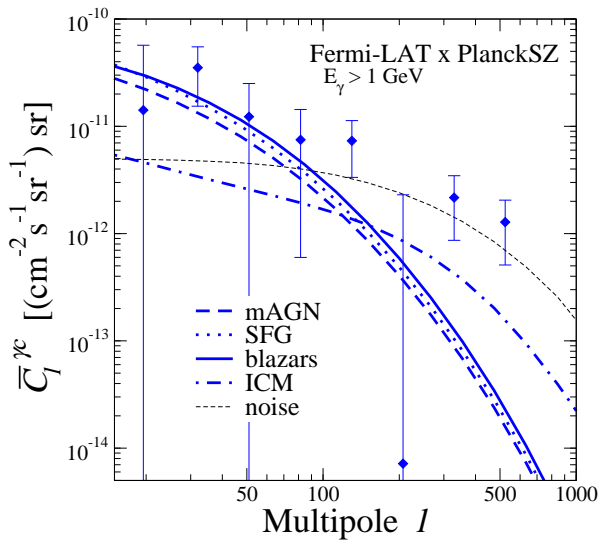}
\caption{CAPS between PlanckSZ clusters and \g-ray emitters (convolved with beam window function $W_\ell^{B}$) compared to the measurement at $E_\gamma>1$ GeV. Dotted line shows a noise term at the level of $C_\ell^{(\gamma c)}=5\cdot10^{-12}{\rm cm^{-2}s^{-1}}$.}
\label{fig:CAPS_Planck}
\end{figure}

The results of the comparisons between measured and predicted cross-correlation between \g-ray emitters   (blazars, mAGN, SFG and ICM) and clusters
 (redMaPPer, WHL12  and PlanckSZ) are shown in Figs. \ref{fig:CAPS_redMA}, \ref{fig:CAPS_WHL12} and \ref{fig:CAPS_Planck}.
 Dots with errorbars refer to the data whereas the different curves represent predictions obtained using the models described in Section~\ref{sec:models}.
 The latter have been normalized (without changing neither spectral nor spatial shapes) such that blazars, mAGNs and SFGs individually contribute to 100\% of the UGRB above 1 GeV.
 For the ICM case we consider a 1~\% contribution to meet the observational constraints described in Section \ref{sec:wf}.

At $\ell\lesssim100$, the data 
are well fitted by a model in which the \g-ray emission is produced by blazars, mAGNs or SFGs (or a combination of them), provided that 
they contribute to 100~\% of the UGRB. 
On the contrary, the ICM contribution to the cross correlation is highly subdominant, largely because of the 1~\% UGRB contribution constraint.
Note however that the ICM would account for much more than 1~\% of the CAPS (while blazars, mAGNs or SFGs providing 100~\% of the UGRB contribute to a similar fraction, roughly 100~\%, of the CAPS). This is because, when compared to galactic \g-ray emitters, the ICM contribution has a relatively larger non-linear term and its window function have a better overlapping in redshift with the catalogs window functions.
Though small, the amplitude of the ICM CAPS is not the same in all panels. Instead it correlates with the average mass of the clusters in the catalogs.
Indeed, one can notice how it increases going from WHL12 to PlanckSZ (the catalogs with, respectively, the smallest and largest average mass, see Fig.~\ref{fig:cat_prop}b), in particular in comparison with the milder increase expected in the amplitudes associated to the other \g-ray emitters.
In the case of PlanckSZ catalog, that contains massive clusters, the ICM contribution is larger than that of the other astrophysical sources 
at  $\ell \gtrsim100$.
  
 The similarity in the theoretical CAPS of all the above \g-ray emitters, except the ICM, guarantees the fact that all of them provide a good fit to the data
 and reflects the similarities in the \g-ray window functions. The small differences originates from two effects: 
 the (relatively small) differences in the shapes of the \g-ray window functions (see, e.g., the Supplemental Material in \citep{Regis:2015zka}) and the different redshift dependences of the bias factors of the various emitters (see, e.g., Appendix of \citep{2015ApJS..221...29C}).

At  $\ell\gtrsim100$ the models systematically underestimate the measured CAPS. As we have discussed,  the validity of our model is expected to break 
down at small scale. For this reason we included a shot-noise-like term that restores the missing power at high $\ell$. This term, convolved 
 with the {\it Fermi}-LAT beam window function, is shown with black dot-dashed lines in Figs. \ref{fig:CAPS_redMA}, \ref{fig:CAPS_WHL12} 
 and \ref{fig:CAPS_Planck}. Its introduction clearly improves the quality of the model that
 now fits the data all the way out to $\ell=1000$.

In principle, one could account for this term by increasing the normalization of the ICM window function by one order of magnitude. Indeed, the shape of the CAPS ICM is similar to the signal shape over the whole multiple range. On the other hand, this increase would make the ICM contribution to the UGRB to grow to $\sim$10\%, something which conflicts with current bounds.

The different \g-ray emitters have different energy spectra, and we can have a
better insight on the origin of the cross-correlation signal by repeating the correlation analysis in finer energy bins. We considered 
eight of them: $[0.25,0.5] \; [0.5,1.0] \; [1.0,2.0] \; [2.0,5.0] \; [5.0,10.0] \; [10.0,50.0] $ $\; [50.0,200.0] \; [200.0,500.0] $ GeV.
Building upon the results of the previous analysis we now move on from the specific benchmark models adopted above and consider a simpler CAPS model:
\be
C_{\ell,a}^{(\gamma c)}=C^{1h}_a+A^{2h}_a C_{\ell,a}^{2h} \;,
\label{eq:simpmodel}
\ee
with $a=1,..,8$ running over the energy bins\footnote{In the correlation with the PlanckSZ catalog, we discard the energy bins 1 and 8 since they are meaningless due to their low statistics.}, $C^{1h}_a$ is the parameter of the fit related to the 1-halo contribution,   $C_{\ell,a}^{2h}$ is the 2-halo model term that we have modeled according to the blazar case since, as we have seen above, the other \g-ray emitters produce a similar cross-correlation signal apart from the ICM that, again according to the previous analysis, is 
subdominant. $A^{2h}_a$ is the second free parameter of the fit and sets the amplitude of the 2-halo term.
 We remind the reader that, in the halo model, the CAPS is generically given by the sum of a 1-halo and a 2-halo terms, as described in Section \ref{sec:models}; in Eq.~(\ref{eq:simpmodel}), the 1-halo CAPS is constant (since the source is assumed to be point-like) and the 2-halo CAPS is computed plugging Eq.~(\ref{eq:PSBd2}) into Eq.~(\ref{eq:clfin}).
 
The null hypothesis (no signal) can be discarded at high significance. Indeed, the $\Delta \chi^2$ 
with respect to this model is 43.2 (WHL12), 53.5 (redMaPPer), and 14.4 (PlanckSZ). Considering the number of parameters of the model (16 for WHL12 and redMaPPer and 12 for PlanckSZ) and following Wilks' theorem~\citep{Wilks:1938}, this leads to $p$-values of $2.6\times10^{-4}$ (WHL12), $6.3\times10^{-6}$ (redMaPPer), and 0.27 (PlanckSZ).
Note that these are conservative estimates of the significance of the signals, since one could reduce the degrees of freedom by fixing the energy spectrum of the model (that, as we will see in the following, is in fair agreement with expectations for blazars).

\begin{figure}[t]
\vspace{-3cm}
\centering
\includegraphics[width=0.49\textwidth]{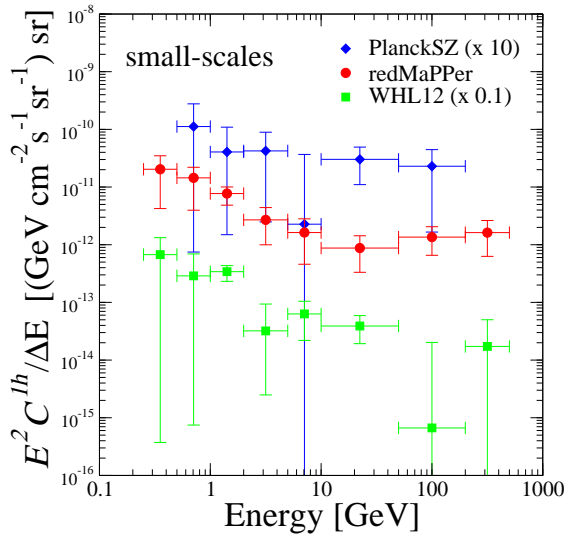}
\caption{Energy spectrum of a constant 1-halo CAPS. This has been derived by fitting the measurements in eight energy bins from 0.25 to 500 GeV with $C^{1h}+A_{2h}C^{2h}_\ell$, where $C^{1h}$ and $A_{2h}$ are the fitting parameters and $C^{2h}_\ell$ is taken to follow the blazar case.}
\label{fig:CAPS_spectrum1h}
\end{figure}

\begin{figure}[t]
\vspace{-2cm}
\centering
\includegraphics[width=0.49\textwidth]{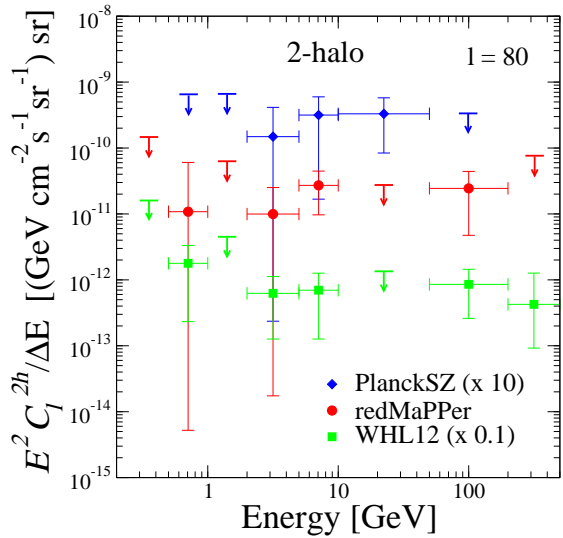}
\caption{Energy spectrum of the 2-halo CAPS term at $\ell=80$. This has been derived by fitting the measurements in eight energy bins from 0.25 to 500 GeV with $C^{1h}+A_{2h}C^{2h}_\ell$, where $C^{1h}$ and $A_{2h}$ are the fitting parameters and $C^{2h}_\ell$ is taken to follow the blazar case. The plot shows $A_{2h}C^{2h}_{80}$.}
\label{fig:CAPS_spectrum2h}
\end{figure}

We show the derived fitting parameters in Figs.~\ref{fig:CAPS_spectrum1h} and \ref{fig:CAPS_spectrum2h}.  In Fig.~\ref{fig:CAPS_spectrum1h}, 
we plot the $C^{1h}$ term for the different energy bins.
The global statistical evidence (i.e. adding up in quadrature the evidences of single energy bins) for the 1-halo component is $3.9\sigma$ (WHL12), $4.7\sigma$ (redMaPPer), and $2.3\sigma$ (PlanckSZ).
Fig.~\ref{fig:CAPS_spectrum2h} instead shows the 2-halo term $A^{2h}C_{\ell=80}^{2h}$ whose statistical evidence 
turns out to be $2.6\sigma$ (WHL12), $2.1\sigma$ (redMaPPer), and $1.8\sigma$ (PlanckSZ).
The evidence in each energy bin for the 1-halo (2-halo) term (and in turn the error bars in Fig.~\ref{fig:CAPS_spectrum1h} (Fig.~\ref{fig:CAPS_spectrum2h})) have been derived by evaluating the likelihood ratio between the model in Eq.~(\ref{eq:simpmodel}) and the same model but with the 1-halo (2-halo) term set to zero.

The energy dependence of the 1-halo term in the redMaPPer case  shows a possible break at $E=10$ GeV which suggests the presence of two contributions:
a hard component with spectral index close to $-2$ above 10 GeV and a softer component at lower energies. The analysis of the other two catalogs 
displays a similar trend. However, the quality of the data does not allow us to draw statistically significant conclusions.
A fit to the energy spectrum of the redMapper 1-halo CAPS with the sum of two power-laws is preferred over the single power-law case at 85\% C.L., with the the best-fit spectral indexes being $-2.9$ and $-2.0$ (while $-2.7$ is the best-fit spectral index in the case of a single power-law).
No break is seen in the energy-dependence of the 2-halo term with a spectral index of $\sim -2$. This is consistent with being  
produced by the same sources responsible for the 1-halo term at $E > 10$ GeV.

Knowing that BL Lac-type of blazar are characterized  by hard energy spectra in contrast with  the softer spectra of 
all other galactic \g-ray emitters (mAGN, FSRQ, SFG) considered in our model, the emerging picture is that of a cross correlation signal 
dominated by BL Lacs on large scale (where the 2-halo term dominates) and on small scales, but possibly only at high energy. 
The soft component, seen in the 1-halo term only, might indicate that the small-scale cross-correlation signal is also contributed 
by a different type of \g-ray emitters that takes over at $E < 10$ GeV. These can be non-BL Lac AGNs or SFGs hosted in the cluster halo, or
the ICM itself (or a combination of them).

\section{Conclusions}
\label{sec:concl}

In this work, we have analyzed the 2-point angular cross-correlation between the unresolved EGB observed by 
{\it Fermi}-LAT and the number of galaxy clusters in three different catalogs: WHL12, redMaPPer and PlanckSZ.
The main results are:
\begin{itemize}

\item
For the first time, we detect a cross-correlation signal in both configuration and Fourier space.
The measurement has a strong statistical significance for WHL12 and redMaPPer, while it is at the level of a hint for PlanckSZ.

\item
An alternative, more conventional, stacking analysis was performed that confirms these results. As expected, the 
\g-ray emission profile that emerges from the stacked images 
turns out to be consistent with that of the angular CCF.

\item
The cross-correlation measurement confirms that the unresolved EGB observed by {\it Fermi}-LAT correlates with the large scale clustering of matter in the Universe (traced by clusters), as found also in \cite{Fornengo:2014cya} and \cite{Xia:2014}, employing different tracers.

\item
We model the cross correlation signal in the framework of the halo model and  
consider \g-ray emission from four types of astrophysical sources: blazars, mAGN, SFG and ICM.
Our model provides a good fit to the measured CAPS on large scales.  At $\ell\lesssim100$
all  these sources but the ICM  generate a cross-correlation compatible with the observations. 
The ICM contribution is subdominant. Its small amplitude is constrained by the fact that no \g-ray emission 
from clusters, including the nearby ones, has been detected so far.
Only in the PlanckSZ cluster catalog case the ICM may provide a non negligible contribution at intermediate scales.

On small scales our model underestimates the observed CAPS.  In this range the angular spectrum is approximately flat and 
can be fitted by a constant, shot-noise-like term, that we included in our model.
Its physical  interpretation is as follows. {\it Fermi}-LAT has a rather large PSF ($\gtrsim0.1^\circ$). 
The recorded arrival direction of photons is ``randomized'' with respect to the true direction on scales below (comparable to) the PSF.
This creates a shot-noise-like term and prevents to fully characterize the correlation among sources beyond the PSF angular scale.
Noting that the mean cluster redshift in the considered catalogs is $\bar z\sim 0.2-0.4$, 
any population of sources with physical sizes below a few Mpc could contribute to the small-scale CAPS.

\item

The cross correlation signal is significantly detected out to $\sim1$ degree, which is beyond the PSF extension (even though the statistical significance of the 2-halo term is not totally conclusive and amounts to $>2\sigma$).
At the typical redshifts of the clusters in the considered catalogs, $\sim1$ degree
corresponds to a linear scale of $\sim 10$ Mpc. This means that a large fraction of the correlation signal seems to be not physically 
associated to the clusters. Instead, it can be produced by AGNs or SFGs residing in the larger scale structures
that surrounds the high density peaks where clusters reside.

\item

Finally, we have investigated the energy dependence of the cross-correlation signal. It turns out that on large scale, where the 2-halo term
dominates, the signal is contributed by sources with hard energy spectra, consistent with that of the BL Lacs.
On small scales, where the 1-halo term dominates, the correlation signal could be contributed by different types of sources.
At high ($E>10$ GeV) energies the dominant sources have hard spectra, i.e. they are probably the same BL Lac population.
At smaller energies, the correlation signal shows a hint of contribution by sources with softer spectra. These can be non-BL Lac AGNs, SFGs 
and/or the ICM.

 \end{itemize}
 
In conclusion, our measurement combined with theoretical expectations suggests that the detected cross-correlation signal is largely contributed by compact sources like AGNs or SFGs. A possible contribution from the ICM, associated to the cluster itself, is however not excluded at small scales. Since its amplitude is expected to increase with the mass of the clusters, a cross-correlation analysis with a wide-field catalog containing a large number of nearby massive clusters can be therefore a suitable way to attempt the detection of the so far elusive \g-ray emission from the ICM.

\section*{Acknowledgements}

The \textit{Fermi} LAT Collaboration acknowledges generous ongoing support
from a number of agencies and institutes that have supported both the
development and the operation of the LAT as well as scientific data analysis.
These include the National Aeronautics and Space Administration and the
Department of Energy in the United States, the Commissariat \`a l'Energie Atomique
and the Centre National de la Recherche Scientifique / Institut National de Physique
Nucl\'eaire et de Physique des Particules in France, the Agenzia Spaziale Italiana
and the Istituto Nazionale di Fisica Nucleare in Italy, the Ministry of Education,
Culture, Sports, Science and Technology (MEXT), High Energy Accelerator Research
Organization (KEK) and Japan Aerospace Exploration Agency (JAXA) in Japan, and
the K.~A.~Wallenberg Foundation, the Swedish Research Council and the
Swedish National Space Board in Sweden.
 
Additional support for science analysis during the operations phase is gratefully acknowledged from the Istituto Nazionale di Astrofisica in Italy and the Centre National d'\'Etudes Spatiales in France.

We would like to thank F. Zandanel for discussions.
This work is supported by the research grant {\sl Theoretical Astroparticle Physics} number 2012CPPYP7 under the program PRIN 2012 funded by the Ministero dell'Istruzione, Universit\`a e della Ricerca (MIUR); by the research grants {\sl TAsP (Theoretical Astroparticle Physics)} and {\sl Fermi} funded by the Istituto Nazionale di Fisica Nucleare (INFN); by the {\sl Excellent Young PI Grant: The Particle Dark-matter Quest in the Extragalactic Sky}.
MV and EB are supported by PRIN MIUR and IS PD51 INDARK grants. MV is also supported by ERC-StG cosmoIGM, PRIN INAF.
JX is supported by the National Youth Thousand Talents Program, the National Science Foundation of China under Grant No. 11422323, and the Strategic Priority Research Program, {\sl The Emergence of Cosmological Structures} of the Chinese Academy of Sciences, Grant No. XDB09000000.
SC is supported by ERC Starting Grant No. 280127.
Some of the results in this paper have been derived using the
HEALPix\footnote{http://healpix.sourceforge.net/downloads.php} package.

\bibliographystyle{apj}
\bibliography{references}

\appendix
\section{Validation and cross-checks}
\label{sec:app1}
\begin{figure*}[t]
\vspace{-2cm}
\centering
\includegraphics[width=0.33\textwidth]{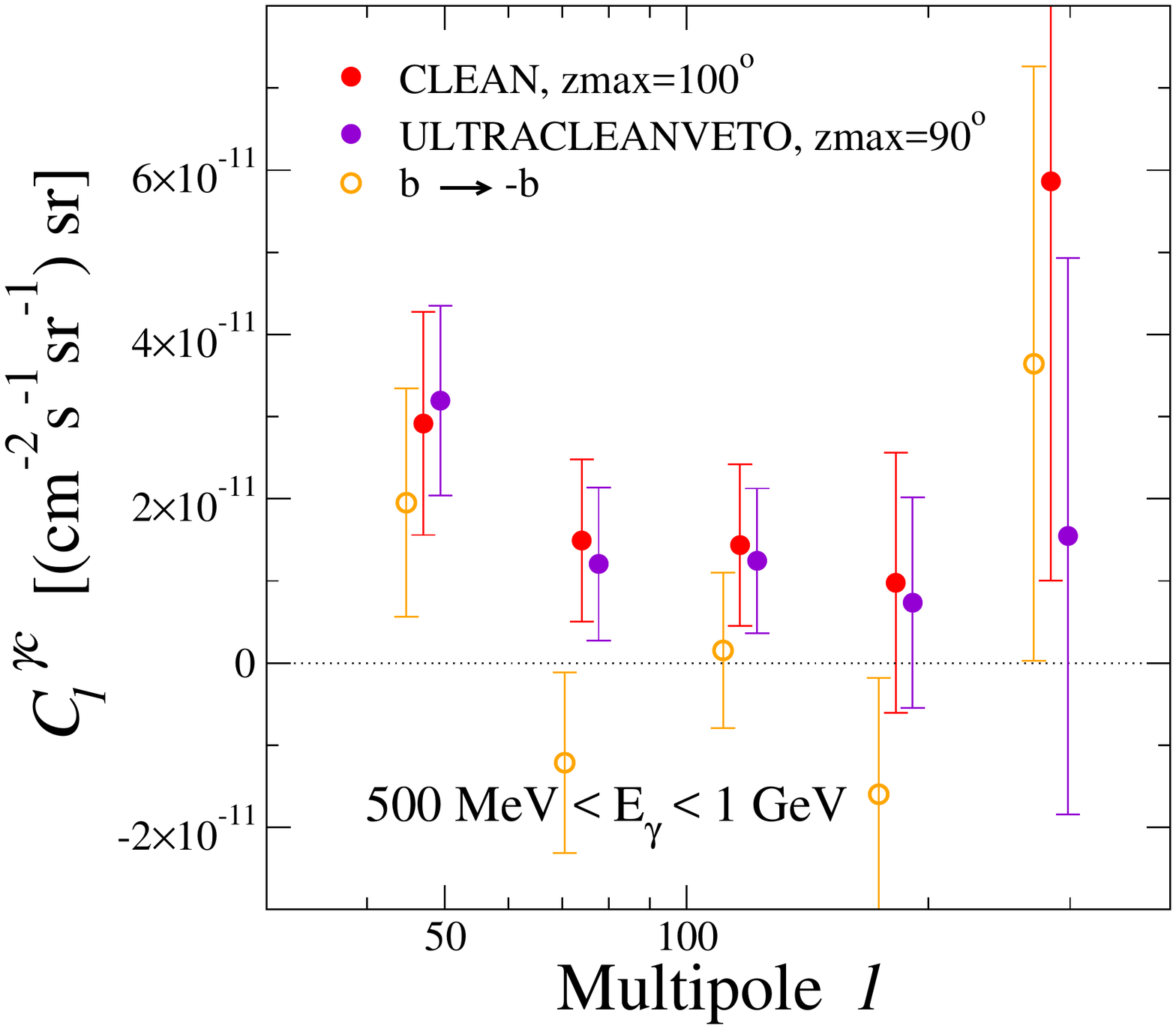}
\includegraphics[width=0.33\textwidth]{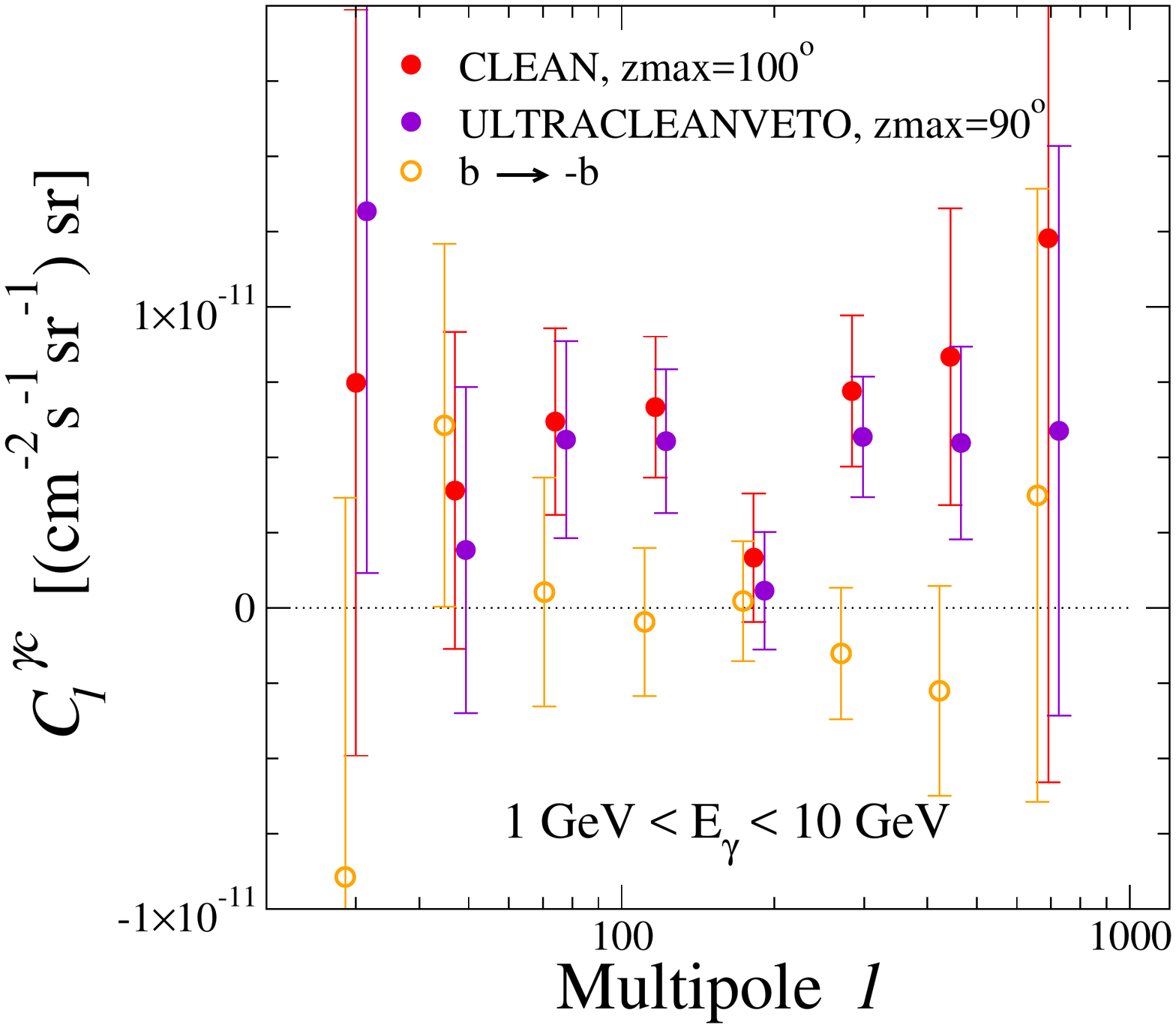}
\includegraphics[width=0.33\textwidth]{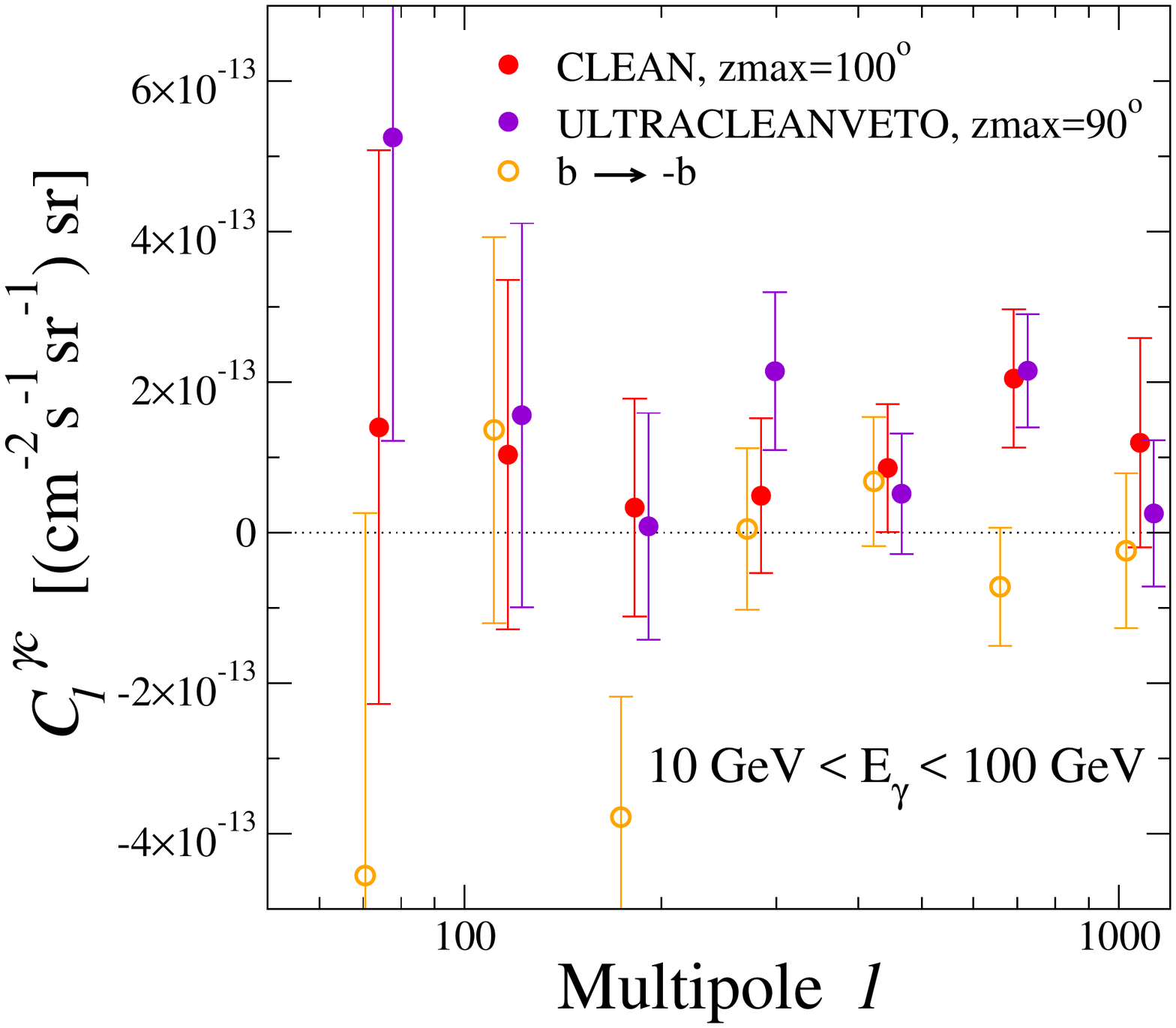}
\caption{Observed CAPS (PSF deconvolved) between the {\it Fermi}-LAT \g-ray map in three different energy bins and the redMaPPer catalog. 
Filled red points show the measurement with the \texttt{P8R2\_CLEAN} {\it Fermi}-LAT event class and zenith angle cut at 100$^\circ$.
Filled violet points show instead the measurement with the \texttt{P8R2\_ULTRACLEANVETO} {\it Fermi}-LAT event class and zenith angle cut at 90$^\circ$.
Open orange points show the CAPS between the {\it Fermi}-LAT \g-ray maps and a mock realization of the catalog where we perform the transformation $b\rightarrow-b$ for each redMaPPer cluster.
Left panel refers to $500\,{\rm MeV}<E_\gamma<1\,{\rm GeV}$, central panel to $1\,{\rm GeV}<E_\gamma<10\,{\rm GeV}$, and right panel to $10\,{\rm GeV}<E_\gamma<100\,{\rm GeV}$.}
\label{fig:CAPS_valid}
\end{figure*}

To assess the robustness of our analyses and results, we performed a few different tests, along the lines described in Section 6 of \cite{Xia:2014}.
No unexpected behavior has been found.
In this Appendix, we report the results for two tests and, for the sake of definiteness, we focus on the redMaPPer catalog only.
First, we considered a different selection of \g-ray events. The analysis in the main text has been conducted using the {\verb"P8R2_CLEAN"} class of {\it Fermi}-LAT events.
This is a common choice for diffuse studies since it provides a good compromise between having a clean event class and sufficiently high statistics. 
Here we analyze also the {\verb"P8R2_ULTRACLEANVETO"} photons, namely the cleanest Pass 8 event class.
Moreover, we considered a more conservative zenith angle cut, excluding photons detected with measured zenith angle larger than 90$^{\circ}$ (instead of 100$^{\circ}$, as in the analysis presented in the main text).
The comparison is shown in Fig.~\ref{fig:CAPS_valid} (red versus violet points). It is clear that the results are fully compatible.
Also the significances of detection are not dramatically affected. For the more conservative choice of photon events, the $p$-values of the statistical analysis described in Section~\ref{sec:res} would become $8.8\times 10^{-3}$ (WHL12), $3.7\times 10^{-5}$ (redMaPPer), and 0.18 (PlanckSZ).

To test the robustness of the detection we built a mock realization of the redMaPPer catalog by performing the transformation on the Galactic latitude $b\rightarrow-b$ for each cluster of the sample. 
This realization preserves the intrinsic clustering of the catalog (i.e., it provides the same autocorrelation signal), but should remove the cross-correlation (for more details, see Section 6.6 in \cite{Xia:2014}).
Indeed, as clear also from Fig.~\ref{fig:CAPS_valid} (orange open points), the derived CAPS is compatible with no signal,
with the $p$-value now becoming $0.994$, meaning no preference over the null hypothesis.

\section{Consistency among different samples of clusters}
\label{sec:app2}
The WHL12 catalog is, among the three samples we considered, the one with the largest number of clusters.
This is due to the fact that it has the lowest richness (and in turn mass) threshold, see also Fig.~\ref{fig:cat_prop}b.
The redshift range is instead not dramatically different, especially between WHL12 and redMaPPer (that also share the fraction of sky probed).

\begin{figure*}[t]
\vspace{-2cm}
\centering
\includegraphics[width=0.49\textwidth]{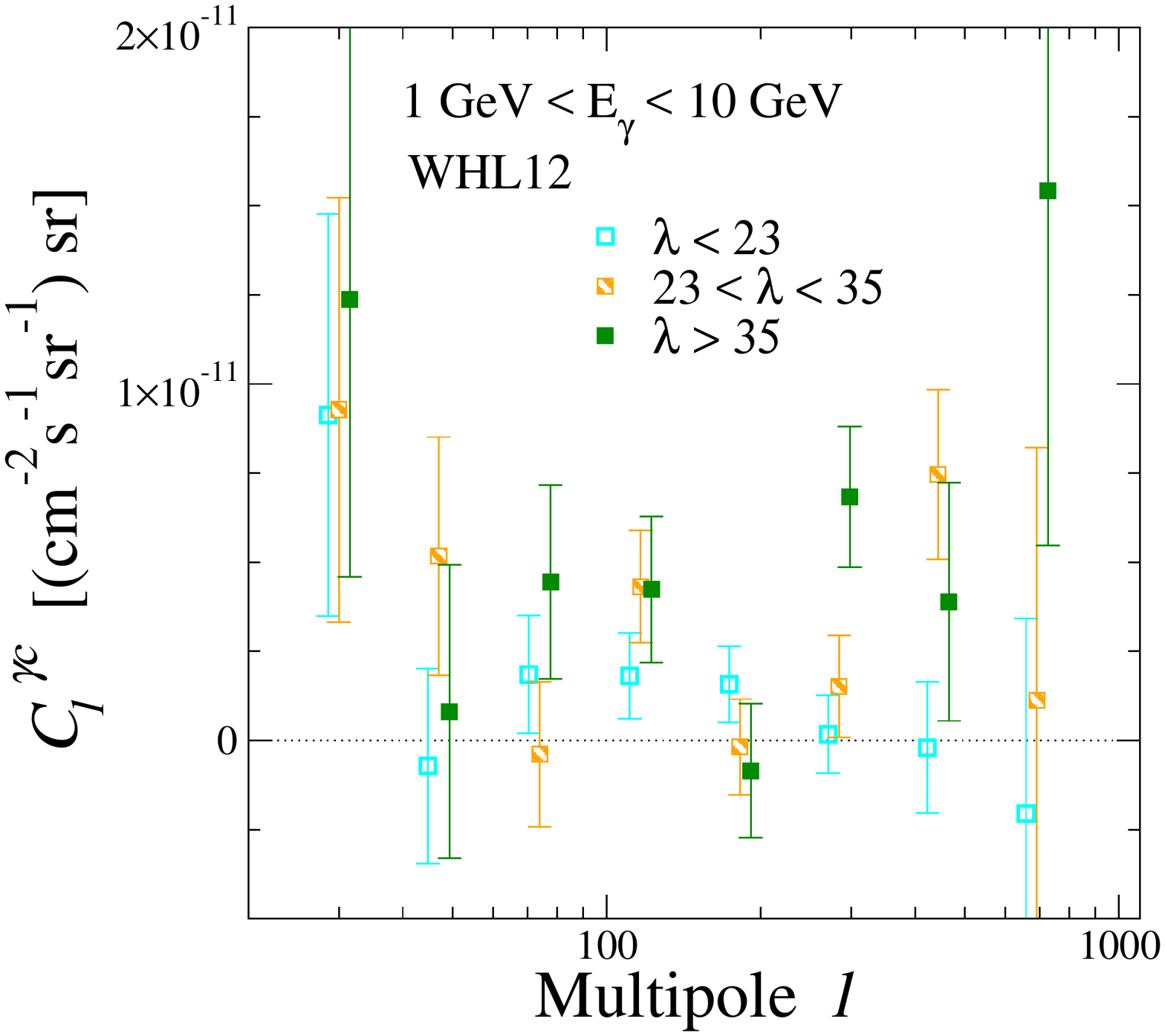}
\includegraphics[width=0.49\textwidth]{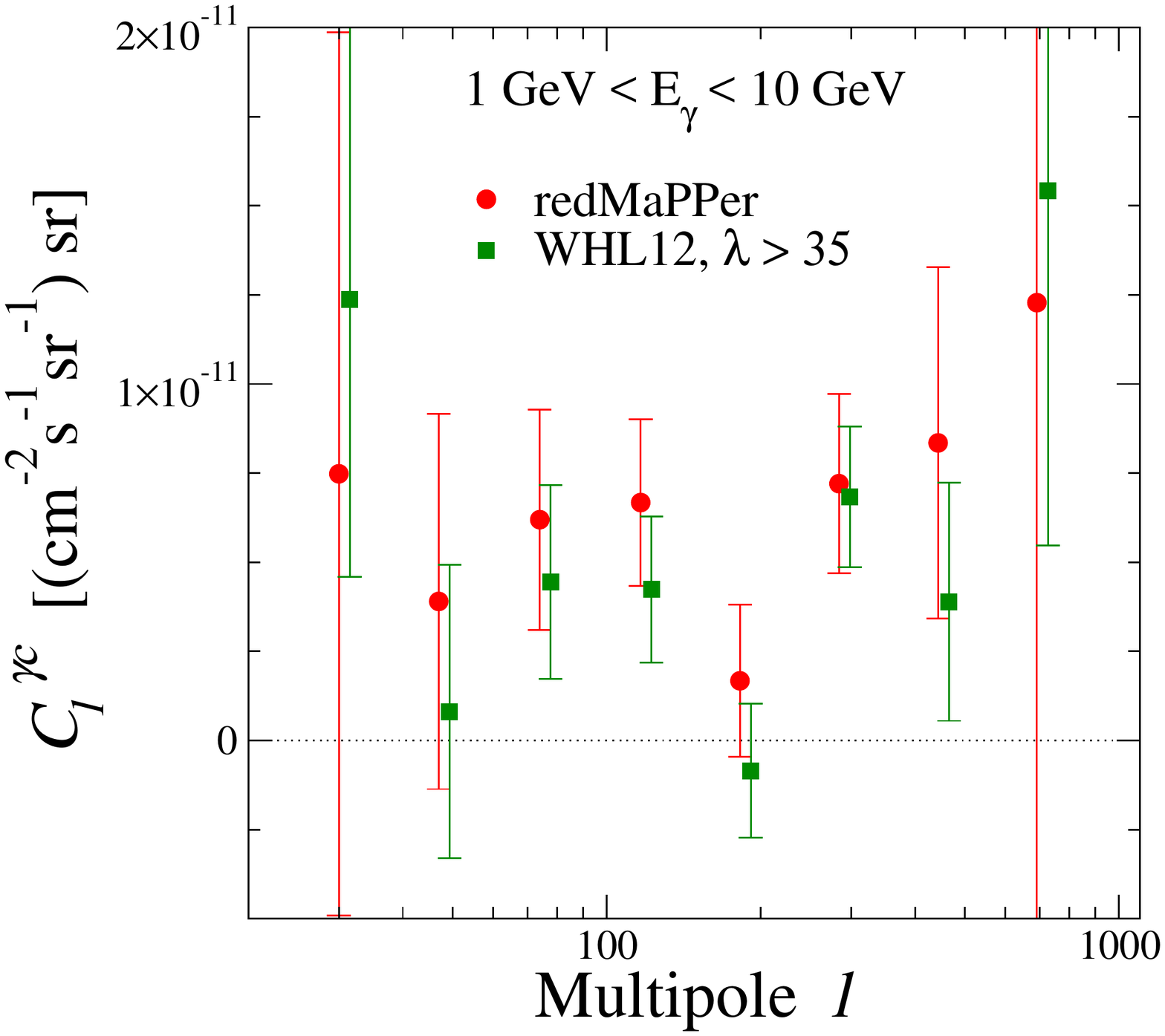}
\caption{Left panel: Observed CAPS (PSF deconvolved) between the {\it Fermi}-LAT \g-ray map at $1\,{\rm GeV}<E_\gamma<10\,{\rm GeV}$ and the WHL12 catalog of clusters for three different richness bins: $\lambda<23$ (cyan), $23<\lambda<35$ (orange) and $\lambda>35$ (dark green). Right panel: Comparison between the CAPS obtained from the cross-correlation with redMaPPer (red) and the clusters with $\lambda>35$ in WHL12 (dark green).}
\label{fig:CAPS_richness}
\end{figure*}

In this Appendix, we test the dependency of the WHL12 cross-correlation signal from the cluster richness (that can be translated into cluster mass, see discussion in Section~\ref{sec:wf}). We also check the consistency of our findings for the different catalogs, by comparing the signal of a high-richness subsample of WHL12 with the redMaPPer case. 
To this aim we split the WHL12 into three samples of richness $\lambda<23$, $23<\lambda<35$ and $\lambda>35$. The last bin contains 24,903 clusters, similarly to the redMaPPer catalog.
Results are shown in Fig.~\ref{fig:CAPS_richness}, where we focus on the \g-ray energy bin $1\,{\rm GeV}<E_\gamma<10\,{\rm GeV}$ for definiteness. Even though the statistics is not very high, left panel shows that the amplitude increases with richness, as expected. The right panel compares the CAPS of the high-richness subsample of WHL12 with the redMaPPer one. The two cross-correlation measurements are fully compatible.

\end{document}